\documentclass[11pt,tightenlines,eqsecnum,floats,aps,amsmath,amssymb,nofootinbib,prd,shownopacs,floatfix,superscriptaddress]{revtex4-1}

\usepackage{graphicx}
\usepackage{epstopdf}
\usepackage{latexsym}
\usepackage{amssymb}
\usepackage{amsmath}
\usepackage{color}
\usepackage{mathrsfs}
\usepackage{xparse}
\usepackage{float}
\usepackage{mathtools}
\usepackage{natbib}

\usepackage[center]{subfigure}

\begin{document}

  \renewcommand\arraystretch{2}
 \newcommand{\bq}{\begin{equation}}
 \newcommand{\eq}{\end{equation}}
 \newcommand{\bqn}{\begin{eqnarray}}
 \newcommand{\eqn}{\end{eqnarray}}
 \newcommand{\nb}{\nonumber}
 \newcommand{\lb}{\label}
 \newcommand{\cb}{\color{blue}}
    \newcommand{\cc}{\color{cyan}}
        \newcommand{\cm}{\color{magenta}}
\newcommand{\rc}{\rho^{\scriptscriptstyle{\mathrm{I}}}_c}
\newcommand{\rd}{\rho^{\scriptscriptstyle{\mathrm{II}}}_c} 
\NewDocumentCommand{\evalat}{sO{\big}mm}{%
  \IfBooleanTF{#1}
   {\mleft. #3 \mright|_{#4}}
   {#3#2|_{#4}}%
}
\newcommand{\PRL}{Phys. Rev. Lett.}
\newcommand{\PL}{Phys. Lett.}
\newcommand{\PR}{Phys. Rev.}
\newcommand{\CQG}{Class. Quantum Grav.}
\newcommand{\parallelsum}{\mathbin{\!/\mkern-5mu/\!}}

\title{Mukhanov-Sasaki  equation in manifestly gauge-invariant linearized cosmological perturbation theory with dust reference fields}
\author{Kristina Giesel}
\email{kristina.giesel@gravity.fau.de}
\affiliation{Institute for Quantum Gravity, Department of Physics, FAU Erlangen-N\"urnberg, Staudtstr. 7, 91058 Erlangen, Germany}
\author{Laura Herold}
\email{lherold@mpa-garching.mpg.de}
\affiliation{Max Planck Institute for Astrophysics, Karl-Schwarzschild-Str. 1, 85748 Garching, Germany}
\author{Bao-Fei Li}
\email{baofeili1@lsu.edu}
\author{Parampreet Singh}
\email{psingh@lsu.edu}
\affiliation{Department of Physics and Astronomy, Louisiana State University, 
Baton Rouge, Louisiana 70803, USA}

\begin{abstract}
The aim of this article is to understand the role of dust reference fields, often also called clocks, on cosmological perturbations around a classical spatially-flat Friedmann-Lema\^{i}tre-Robertson-Walker (FLRW) universe. We derive the Mukhanov-Sasaki equation for the Brown-Kucha\v{r} and Gaussian dust models, which both consider four dust fields as reference fields. The reduced phase space of Dirac observables, that is the gauge-invariant part of the theory,  is constructed by means of an observable map applied to all elementary phase space variables of the coupled system, consisting of gravity, a massive scalar field and the dust degrees of freedom and automatically yields the set of independent physical variables. The evolution of these observables is governed by a so called physical Hamiltonian which can be derived once the set of reference fields are chosen and differs for each model. First, the reduced phase space for full general relativity as well as the corresponding equations of motion are derived for full general relativity. Then from this, the gauge-invariant versions of the equations of motion for the background are derived which contain a fingerprint of the dust reference fields. Afterwards we study linear cosmological perturbations around a FLRW metric using the scalar-vector-tensor decomposition and derive the equation of motion for the Mukhanov-Sasaki  variable in this formalism for a chosen set of variables on the reduced phase space and expressed in terms of Dirac observables. 
The Mukhanov-Sasaki equation involves additional contributions that can be understood as back reactions from the dust reference fields. These additional dust contributions to the Mukhanov-Sasaki equation were absent if the dust energy and momentum density as well as their perturbations are vanishing. 
The nature of the correction terms suggests that Brown-Kucha\v{r} and Gaussian dust reference fields contribute differently.  We numerically study the behavior of the dust contributions to the  Mukhanov-Sasaki equation during inflation.
\end{abstract}
\maketitle

\section{Introduction}
\label{Intro}
\renewcommand{\theequation}{1.\arabic{equation}}\setcounter{equation}{0}
General relativity (GR) can be understood as a gauge theory where the role of gauge symmetries is played by diffeomorphisms. Since this gauge group is more complicated than for other gauge theories in physics, the construction of corresponding gauge-invariant quantities so called Dirac observables is also a non-trivial task. At the level of canonical GR Dirac observables are required to Poisson commute with the first class constraints that generate diffeomorphisms on the ADM phase space. A framework in which such observables  can be constructed systematically is the relational formalism. This formalism was first studied  by Bergmann, Komar and Kucha\v r \cite{bergman1961,bergman2, komar1958, kuchar1991} and then conceptually improved by Rovelli \cite{rovelli, rovelli2}. Its mathematical structure was analyzed in \cite{Vy1994} and then further developed by Dittrich \cite{dittrich, dittrich2} and Thiemann \cite{thiemann2006}. The idea of this formalism is to introduce so called reference fields that provide a physical reference system with respect to which the dynamics of the remaining degrees of freedom in a given system is formulated. A primary advantage of this relational picture is that it can be used to obtain for a given function on phase space its corresponding gauge-invariant extension and hence Dirac observable. Moreover, it provides a conceptually clear way to formulate dynamics of Dirac observables in a relational manner and thus avoids the problem of time often referred to in the context of GR. The observable map introduced in \cite{Vy1994, dittrich} allows to derive the reduced phase space of GR, that is the phase space of only the independent physical degrees of freedom if suitable reference fields are chosen. The class of types of reference fields for which this has been done so far are mainly either dust scalar fields or Klein-Gordon scalar fields, see for instance \cite{gt2015} for an overview and references therein. More exotic reference fields can for instance be found in \cite{thiemann2006,gv2017}. Given a specific choice of reference fields the generator for the dynamics of the observables -- the physical Hamiltonian -- can be derived. In general it  will differ for each given model and also provide a  finger print of the reference fields in the final gauge-invariant equations of motion. This formalism has been applied to various settings in GR \cite{dittrich2,dt2006,ghtw2010I,ghtw2010II,gt2015,Ali:2015ftw,gh2018,ghs2018,gsw2019,laura}, scalar-tensor theories \cite{hgm2015}, LTB spacetimes \cite{gtt2010} and  loop quantum gravity \cite{gt2007,dgkl2010,hp2012,hp2012,gv2016,gv2017,Ali:2018vmt,hl2019I,hl2019II}. Some of the reasons why these kinds of reference fields have been chosen in the past is that they ensure that the physical Hamiltonian is a constant of motion and the algebra of observables has a very simple structure. The latter is in particular important once quantization of gravity comes into play.

However, even within this class of types of reference fields an important question in the relational formalism is the way reference fields may impact physical predictions. By this we mean that although everything is formulated at the gauge-invariant level,  choosing different reference fields in the relational formalism corresponds to coupling different additional matter fields to GR which is also coupled to other non-reference matter fields. Thus, each model with a specific choice of reference fields can in principle have characteristic features resulting from this choice. In this work, we consider this question in the context of cosmological perturbation theory where the construction of gauge-invariant quantities plays a pivotal role. The idea that reference fields must not play any role except of test fields in for instance cosmological dynamics is only an idealization that no longer holds if we work in the relational formalism where the reference fields are actually coupled as dynamical degrees of freedom to the system under consideration. Given that the energy density of reference fields in general evolves at a different rate than other matter components,  there can be a situation in which even when starting with small energy densities compared to other matter fields, cosmic evolution results in a  significant increase of the reference field density. As an example, consider a case of a cosmological spacetime sourced with a massless scalar field or stiff matter along with dust reference fields. The energy density of the latter decays as inverse of the volume of the universe, while for a massless scalar field energy density decays as inverse of volume squared. Thus, even if one starts with an energy density of the dust reference fields which is much smaller than the initial energy density of the massless scalar field, due to expansion of the scale factor the former becomes larger than the latter after a certain time. However, such a situation can be avoided in the case of an inflationary field where one can choose appropriate initial conditions for the dust reference fields such that they behave as approximately test fields in the entire evolution. In this case certain sets of initial values for the densities of the reference fields can be ruled out easily by numerically studying the background evolution and by demanding suitable initial conditions such that sufficient inflationary e-foldings occur. Apart from effects in the background dynamics, more intricate and subtle effects due to reference fields can occur in cosmological perturbations. Here the issue is the way reference fields affect the evolution equations for cosmological perturbations such as for instance the Mukhanov-Sasaki variable. 
Whether or not the correction due to the reference fields generate an appreciable effect in cosmological perturbations, and how the choice of different reference fields can result in distinct effects is a pertinent issue which we aim to explore in this work.  

In this article, we are interested in understanding the role of the dust reference fields in linearized cosmological perturbation theory. We consider GR minimally coupled to a massive scalar field (in a Starobinsky potential) and coupled  in addition to dust reference fields which act as observers in the framework of the relational formalism. We consider both the Gaussian and Brown-Kucha\v{r} dust as reference fields. The coupling of matter reference fields allows to construct the manifestly gauge-invariant quantities at the level of full GR before any perturbations are considered as it has for instance been done in \cite{ghtw2010I,ghtw2010II,gtt2010,hgm2015} and hence obtain the reduced phase space for the full general relativistic setup. An advantage to use these dust reference fields compared to geometrical clocks, as it has been done in \cite{gh2018,ghs2018,gsw2019}, is that the resulting  Hamiltonian constraint in the extended ADM phase space can be easily expressed in deparametrized form at the level of full GR. As a result, the  physical Hamiltonian in the reduced phase space that generates the dynamics of the observables can be easily derived. Once Hamilton's equations are available, one can proceed to consider cosmological perturbations and one can study their impact on the primordial scalar power spectrum via the modified Mukhanov-Sasaki equation in the relational formalism with dust reference fields.  Unlike in conventional cosmological perturbation theory in which the gauge-invariant perturbed quantities need to be constructed order by order, cosmological perturbation theory in the relational formalism is formulated in the reduced phase space which consists only of gauge-invariant observables from the very beginning. For  a presentation of the conventional gauges used in linearized cosmological perturbation theory formulated in the framework of the relational formalism and Dirac observables, see \cite{gh2018,ghs2018,gsw2019}. For a general setup of a perturbative approach to Dirac observables, see \cite{dt2006} and for an application to cosmology \cite{dt2007}. In this approach, in the case of the aforementioned matter reference fields  all observables are invariant under finite gauge transformations and one has no corrections in the gauge invariance condition that are higher than the order in perturbation theory that one considers.  As a result, the discussion of the higher-order perturbations in the relational formalism is more straightforward and transparent to all orders.

The Gaussian and Brown-Kucha\v{r} dust fields introduce four additional degrees of freedom when coupled with gravity. Since the number of constraints stays the same, this results finally in four additional physical degrees of freedom in the reduced phase space independently of the choice of independent variables that we choose to describe the reduced phase space. These additional degrees of freedom leave their imprints in the equations of motion for perturbations. In a scalar-vector-tensor (SVT) decomposition for the linearized perturbations, these extra degrees of freedom manifest themselves in two additional degrees of freedom in the scalar sector and two in the vector sector. Due to these additional degrees of freedom, the choice of dust reference fields can in principle produce detectable deviations from the conventional approach  in cosmological perturbations where usually geometric reference fields are used. In this work, we derive the modified Mukanov-Sasaki equation with Gaussian and Brown-Kucha\v{r} dust fields  as reference fields. More specifically,  on the reduced phase space chosen a set of independent variables, we find the equation of motion for the image of the  Mukhanov-Sasaki variable (defined via eq.(\ref{4b8})) in the standard cosmology theory using the observable map. We then compare the modified Mukhanov-Sasaki equations for above generalization of the Mukhanov-Sasaki variable in the Gaussian and Brown-Kucha\v{r} models by analyzing the correction terms arising from dust reference fields in both of the models. In order to understand the role of different reference fields, as a first step, we estimate the effect of the correction terms in the Mukhanov-Sasaki equation. In particular, we study the way coefficients of different perturbations evolve during inflation. We find these background coefficient terms to decay rapidly with expansion of the universe, especially during inflation. For the reason that the Brown-Kucha\v{r} dust model was originally derived with a negative dust energy density \cite{ghtw2010I}, we consider both cases when dust energy density takes either positive or negative sign in our analysis. 

This paper is organized as follows. In Sec. \ref{sec:Review}, we begin with a brief summary of the relational formalism in GR when gravity is coupled to a single massive scalar field. The construction of the observable map is discussed in a concise way in the extended ADM phase space where lapse and shift are also dynamical variables. Then, we choose  Gaussian and Brown-Kucha\v{r} dust fields as reference fields, construct the reduced phase space with respect to the dust fields and find the corresponding physical Hamiltonians relying on earlier work in \cite{ghtw2010I,gt2015}. We discuss equations of motion for background variables using dust reference fields in Sec. \ref{sec:BackGround}. In this section we also present numerical solutions of equations of motion for some representative cases for the Starobinsky inflationary potential, including when the dust energy density is chosen to be negative, a case which is allowed in Brown-Kucha\v{r} formalism. We find that dust energy density can change the number of inflationary e-foldings. In Sec. \ref{sec:Perturbation}, starting from Hamilton's equations at the gauge-invariant level, we derive  the equations of motion of the linear perturbations around a spatially-flat FLRW universe. Then, with the help of a SVT decomposition choosing a set of independent variables on the reduced phase space, the equations of motion of the scalar modes  and the image of the Mukhanov-Sasaki variable under the observable map defined using dust reference fields are derived. From these equations we find correction terms originating from dust reference fields for Brown-Kucha\v{r} and Gaussian dust models. Our analysis shows that even though the Mukhanov-Sasaki equation takes the same form in the Gaussian and the Brown-Kucha\v{r} dust models, there is a difference in the evolution of the Mukhanov-Sasaki variable because of the different behavior of the  perturbed dust energy density  $\delta \mathcal E^\mathrm{dust}$. In our analysis, this difference is confirmed numerically by solving the perturbed Hamilton's equations. Solving background dynamics numerically, we investigate the evolution of coefficient terms consisting of background quantities of different perturbations in the back-reaction terms. Our analysis shows that all these terms decay rapidly  during inflation. We summarize our main results in section \ref{sec:Conclusion}.

In our paper, we will use $\hbar=c=1$ while keeping Newton's constant $G$ explicit in equations. For numerical studies, Newton's constant is also set to unity. Greek letters are used to denote the 4-dimensional spacetime indices while the Latin letters $a,b,c\dots$ are for the indices of  the tensors on the 3-dimensional hypersurface. 
 
\section{Review of the relational formalism with dust reference fields}
\label{sec:Review}
\renewcommand{\theequation}{2.\arabic{equation}}\setcounter{equation}{0}

In this section, we will first give a brief overview of the relational formalism  in GR considering generic reference fields and then specialize the formalism to the case where dust fields are chosen as reference fields. In particular, we discuss the Gaussian and the Brown-Kucha\v{r} dust field models.  As the relational formalism  and also dust reference fields  have been extensively studied in the literature \cite{bergman1961, bergman2, rovelli, komar1958, kuchar1991, dittrich, gh2018, ghs2018, thiemann2006, gt2015, ghtw2010I, ghtw2010II,laura, Vy1994, g2008, rovelli2, dittrich2}, we will sketch the basic ideas on how to derive the physical Hamiltonian that generates the dynamics on the reduced phase space in each of the dust models and also  give the resulting Hamilton's equations. Finally,  we will end up this section with a comparison  between the two dust reference frames. 

\subsection{The relational formalism in the context of dust reference fields}
\label{sec:RelForm}

The main idea in the relation formalism is that the value of any field at one particular spacetime coordinate does not have any physical meaning due to the diffeomorphism invariance of GR.  Instead, the real physical observable is the value of one field when another field (the reference field) takes some particular value. Furthermore, in order to describe the dynamics of any observables in GR, one also has to choose an appropriate temporal reference field, with respect to which the evolution of the system is unfolded. In principle, these reference fields can be chosen either from some additional matter degrees of freedom, like scalar or dust fields  as for instance done in \cite{ghtw2010I,gtt2010,t2006}, or from the geometrical degrees of freedom of GR itself (see \cite{ghs2018,gsw2019,gh2018}). Although the latter ansatz avoids introduction of extra degrees of freedom into the system, it is more  complicated to formulate the Hamiltonian constraint in deparametrized form and thus one also runs into difficulty when trying to extract the physical Hamiltonian for the reduced system. As a result, in the following,  we will  focus on the first ansatz in which the dust fields are introduced as reference fields. This allows to easily  derive a well-defined physical Hamiltonian. The only price to pay is the appearance of some additional gauge-invariant degrees of freedom. This is  because, as compared to the conventional approach, we consider  four additional degrees of freedom from the beginning pertaining to dust reference fields. Moreover, a single massive scalar field will also be included to source the inflationary phase at late times.  In this section,  we will first focus on the relational formalism when gravity is coupled to a scalar field, then proceed with details when the particular dust field models are considered in the framework in the next two subsections.

The system under consideration in our work consists of a generic scalar field $\varphi$ and dust fields  on a four-dimensional globally hyperbolic spacetime $(\mathcal M, g)$. Thus, the total action is given by
\bq
\lb{action}
S=S_\mathrm{geo}+S_\mathrm{scalar}+S_\mathrm{dust},
\eq
with the geometrical  sector  given explicitly  by 
\bq
\lb{geo}
S_\mathrm{geo}=\frac{1}{\kappa}\int_{\mathcal M} d^4x\sqrt{-g} R^{(4)},
\eq
where $\kappa =16\pi G$ and $R^{(4)}$ denotes the four-dimensional Ricci scalar and $g=\det(g)$. In addition, the action of the scalar field is given by 
\bq
\lb{matter}
S_\mathrm{scalar}=\frac{1}{2\lambda_\varphi}\int_{\mathcal M} d^4x \sqrt{-g} \left(-g^{\mu \nu} \partial_\mu\varphi \partial_\nu\varphi-V(\varphi)\right),
\eq
where  $\lambda_\varphi$ is a coupling constant allowing for a dimensionless $\varphi$ and $V(\varphi)$ denotes the potential term of the scalar field.  
In this section, we will consider the relational formalism in the extended ADM phase space of the gravitational and the scalar field sectors for a  generic reference field at the moment. As expected, the specialization to dust fields as reference fields allows to explicitly consider the additional contributions to the Hamiltonian and diffeomorphism constraints since only the total constraint consisting of the contributions of gravity, the scalar field and the dust fields is required to vanish. For the reason that on the one hand coupling the dust puts the original first class system into a second class one and on the other hand we will choose two different dust models as reference fields, first we briefly review the phase space formulation of gravity plus a scalar field only and exclude the dust fields in our discussion in  subsection \ref{sec:ADM}.  Then we present the basics of the construction of observables via an observable  map and the evolution of observables in \ref{sec:Obsmap} for generic reference fields and finally continue with a discussion on the Gaussian and Brown-Kucha\v{r} dust models in \ref{sec:RelDustGD} and \ref{sec:RelDustBK}.

\subsection{ADM formulation for gravity plus a scalar field}
\label{sec:ADM}
It is well-known that starting from the actions (\ref{geo}) and (\ref{matter}), the Hamiltonian of the system can be derived from the ADM formalism, which uses that a globally hyperbolic spacetime $\mathcal M$ has the topology $\mathcal M\simeq \mathbb{R}\times \sigma$, where $\sigma$ is a spatial 3-dimensional manifold, see \cite{geroch} and \cite{BernalSanchez} for a generalization of the Geroch theorem. Given this $\mathcal M$ foliates into spatial hypersurfaces $\Sigma_t=X_t(\sigma)$, where $X_t$ is an embedding of $\sigma$ into $\mathcal M$ for a fixed value of $t\in\mathbb{R}$. Such an embedding can be conveniently parametrized by its corresponding deformation vector field $\frac{\partial X^\mu}{\partial t}=N(X)n^\mu+N^\mu(X)$, here  $n^\mu$ is a normal unit vector to $\Sigma_t$ and $N^\mu$ is tangential, whereas $N$ and $N^\mu$ are called the lapse function and shift vector respectively. The foliation is then used to perform a 3+1 decomposition of the action. Given the embedding, the inverse metric $g^{\mu\nu}$ can be decomposed into a normal and tangential part as $g^{\mu \nu}=-n^\mu n^\nu+q^{\mu\nu}$. Here $q^{\mu\nu}$ is the tangential part that by means of the embedding can be expressed in terms the inverse ADM-metric $q^{ab}$ on $\sigma$ using $q^{\mu\nu}=q^{ab} X^\mu_a X^\nu_b$.
 Moreover, the unit normal vector in the ADM frame is related with lapse and shift via $n^\mu\coloneqq (\frac{1}{N},-\frac{N^a}{N})$ where $N^a=X^a_\mu N^\mu$ involves the inverse of the embedding denoted by $X^a_\mu$. 
 
 As a result, in the Hamiltonian formalism, the extended ADM phase space $\Gamma_\mathrm{ext}$, which is obtained by pulling back the embedded quantities onto the spatial manifold $\sigma$, consists of  22 degrees of freedom, which are lapse  $N$, shift $N^a$, the ADM metric $q_{ab}$ and the scalar field $\varphi$, as well as  their respective conjugate momenta  denoted by $\pi$, $\pi_a$, $p^{ab}$ and $\pi_\varphi$. The only non-vanishing Poisson brackets of these canonical variables are
\bqn
\lb{poisson}
\{q_{ab}(t,\vec x), p^{cd}(t,\vec y)\}&=&\kappa \delta^c_{(a}\delta^d_{b)}\delta^3(\vec x-\vec y),\nb\\
\{\varphi(t,\vec x), \pi_\varphi(t,\vec y)\}&=&\delta^3(\vec x-\vec y),\nb\\
\{N(t,\vec x), \pi(t,\vec y)\}&=&\kappa \delta^3(\vec x-\vec y),\nb\\
\{N^a(t,\vec x),\pi_b(t,\vec y)\}&=&\kappa \delta^a_b \delta^3(\vec x-\vec y).
\eqn
Then,  the resulting Hamiltonian is a linear combination of the primary and secondary constraints which reads \cite{gh2018}
\bq
\lb{2a1}
H_\mathrm{ext}=\frac{1}{\kappa} \int_\sigma d^3x \left(N c +N^a c_a+\lambda \pi +\lambda^a \pi_a\right),
\eq
in which $\lambda$ and $\lambda^i$ are Lagrange multipliers. The Hamiltonian constraint $c$ and the spatial diffeomorphism constraint $c_a$ (collectively denoted by $c_\mu$) consist of two contributions, that is $c_\mu=c_\mu^\mathrm{geo}+c_\mu^\mathrm{scalar}$ with 
\bqn
\lb{2a2}
c^\mathrm{geo}&=&\frac{1}{\sqrt{q}}\left(p_{ab}p^{ab}-\frac{1}{2}p^2\right)-\sqrt{q}R^{(3)},\\
\lb{2a3}
c^\mathrm{scalar}&=&\frac{\kappa}{2}\Bigg\{\frac{\lambda_\varphi}{\sqrt{q}}\pi^2_\varphi+\frac{\sqrt{q}}{\lambda_\varphi}\left(q^{ab}\partial_a \varphi \partial_b \varphi+V\right)\Bigg\},\\
\lb{2a4}
c^\mathrm{geo}_a&=&-2q_{ab}D_cp^{bc},\\
\lb{2a5}
c^\mathrm{scalar}_a&=&\kappa \pi_\varphi \partial_a \varphi.
\eqn
In the Hamiltonian (\ref{2a1}), $\pi$ and $\pi_a$ are four primary first-class constraints while $c$ and $c_a$ are four secondary first-class constraints, yielding for gravity coupled to a massive scalar field a total of 6 physical degrees of freedom in phase space or equivalently 3 physical degrees of freedom in the configuration space. The whole system is fully constrained with vanishing Hamiltonian on the constraint surface. 

Often one starts directly with the reduced ADM phase space where the primary constraints have already been reduced and lapse and shift are treated as Lagrange multipliers. However, this is not possible if we are interested in gauge-invariant versions for the degrees of freedom for the latter. For instance in the longitudinal gauge used in cosmological perturbation theory where one of the Bardeen potentials is associated with the lapse perturbation, see \cite{gh2018,ghs2018}. For this it is necessary to discuss the relational formalism at the level of the extended phase space which allows a setting closer to the conventional approach based on the Lagrangian framework. In this approach, one can naturally express lapse and shift degrees of freedom in terms of the physical gauge-invariant degrees of freedom in the reduced phase space.

\subsection{Observable map and physical evolution in the relational formalism}
\label{sec:Obsmap}
The relational formalism can be formulated in both the reduced ADM phase space and extended  ADM phase space. The latter where lapse and shift are still dynamical degrees of freedom has the advantage that the observable map can also be applied to lapse and shift degrees of freedom and hence is closer to the Lagrangian formulation where diffeomorphisms act on all degrees of freedom of the metric. Let us as before assume that our system consists of gravity and a minimally coupled massive scalar field, then the secondary constraints are $c^{\rm tot}=c$ and $c_a^{\rm tot}=c_ a$. In the next subsection when we introduce the dust reference fields, of course their contributions have to be added to the total constraints. 
Any Dirac observable satisfies the condition that it has to Poisson commute with the first class constraints of the system under consideration, which in the case of GR are the constraints that generate diffeomorphisms on the extended ADM phase space. The generator consists of a linear combination of the primary constraints $\pi,\pi_a$ and secondary constraints $c,c_a$ and is given by
\bq
\label{modifiedgenerator}
G_\bold b=\frac{1}{\kappa}\left(c[b]+c_a[b^a]+\pi[\dot b+b^a\partial_aN-N^a\partial_a b]+\pi_a[\dot b^a+q^{ac}\left(b\partial_c N-N\partial_c b\right)+b^c\partial_j N^a-N^c\partial_c b^a]\right)
\eq
which was first introduced in \cite{pg2000}. Here $b$ is a general smearing function and $b^a$ a generic vector valued smearing function. Hence, we aim at constructing observables $O$ that at least weakly commute with the generator in (\ref{modifiedgenerator}), that is $\{O,G_\bold b\}\approx 0$. Note that for quantities that do not depend on lapse and shift degrees of freedom this requirement reduces to the one that they have to (weakly) Poisson commute with the secondary constraints $c,c_a$, which is exactly the condition one uses in the case of the reduced ADM phase space. The additional terms in $G_\bold b$ only affect the lapse and shift degrees of freedom and are designed in such a way that they generate diffeomorphisms for these variables. 

The relational framework allows to construct for a given phase space function its associated observable, also often called its gauge-invariant extension once a set of reference fields has been chosen. The number of necessary reference fields is chosen by the number of gauge generators which are four in the case of GR. We denote the reference fields by $T^\mu$, $\mu=0,\dots 3$ one for the temporal and three for spatial diffeomorphisms, which we write in compact notation as $c_\mu=(c,c_a)$. Further let us introduce the multi-index $I$ with constraints $c_I=(c_\mu,\pi_\mu)$ and reference fields $T^I=(T^\mu,\dot{T}^\mu)$. As far as the primary constraints are concerned no independent reference fields need to be chosen but the observable formula takes this automatically into account in terms of $\dot{T}^\mu$. Furthermore, the observable map relies on the assumption that the reference fields are at least weakly canonically conjugate to the constraints. This might not be satisfied for a given choice of reference fields but can always be achieved if the reference fields satisfy the following condition at least locally,
\bq
\mathrm{det}(\mathcal A^I_J )\ne 0 \quad \mathrm{with}\quad  \mathcal{A}^I_J(t,\vec x,\vec y)=\frac{1}{\kappa}\{T^I(t,\vec x), c_J(t,\vec y)\}.
\eq
Explicitly the matrix $\mathcal{A}^I_J$ reads
\begin{equation}
	\label{abelianisation_matrix}
	\mathcal{A}^I_J(t,\vec{x},\vec{y}) 
	\simeq \frac{1}{\kappa} \begin{pmatrix}
		\{T^\mu(t,\vec x),c_\nu(t,\vec y)  \}
		& 				0 				 	\\
		\{\dot{T}^\mu(t,\vec x),c_\nu(t,\vec y) \} 
		&   \{T^\mu(t,\vec x),c_\nu(t,\vec y) \} 	 \\
	\end{pmatrix} .
\end{equation}
The requirement that the inverse matrix denoted by $\mathcal B^I_J$ exists boils down to the condition for the reference fields that $\det(\{T^\mu(t,\vec x),c_\nu(t,\vec y)\})\not=0,\quad \mu,\nu=0,\dots,3$. This property can be applied to weakly abelianize the first-class constraints $c_I$ by defining a new equivalent set of constraints $\tilde {c}_I$ via
\bq
\tilde c_I(t,\vec x)=\int d^3y \mathcal B^I_J(t,\vec x, \vec y) c_J(t,\vec y)\quad {\rm satisfying}\quad  
\{T^I(t,\vec x),\tilde c_J(t,\vec y)\}\approx \kappa \delta^\mu_\nu\delta(\vec x-\vec y).
\eq
Given this we now introduce the gauge fixing conditions for the reference fields of the form ${\mathcal G}^I=({\mathcal G}^\mu,\dot{\mathcal G}^\tau)$ with ${\mathcal G}^\mu=\tau^\mu-T^\mu$, where $\tau^\mu$ is for each $\mu$ a function on ${\cal M}$ but not a dynamical variable on phase space. With this we can construct an observable map for a function $f$ that does not depend on the reference field degrees of freedom on the extended ADM phase space, which maps $f$ to its gauge-invariant extension $O_{f,T^\mu}$, that is $f\mapsto {\mathcal O}_{f,T^\mu}(\tau^\mu)$ with 
\begin{equation}
\label{obsmap}
\begin{split}
 {\mathcal O}_{f,T^\mu}(\tau^\mu)
 &= f+\sum\limits_{n=1}^\infty\frac{1}{n!\kappa^n}\prod\limits_{k=1}^n\int\limits_\sigma d^3x_k{\mathcal G}^J(x_k)\{f(x),\tilde{c}_J(x_k)\}_{(n)},
\end{split}    
\end{equation}
where $\{f,g\}_{(n)}$ denotes the iterated Poisson bracket with 
\bq
\{f,g\}_{(0)}=f ~~~\mathrm{and}~ ~~\{f,g\}_{(n)}=\{\{f,g\}_{(n-1)},g\}.
\eq
$\mathcal O_{f, T}(\tau)$ is the image of $f$ under the  observable map. It is the value of $f$ in the gauge in which the reference fields $T^\mu$ take the values of $\tau^\mu$. As shown in \cite{dittrich,thiemann2006,pg2000} the observable $\mathcal O_{f, T}(\tau)$ commutes with all constraints by construction and is indeed a Dirac observable. This observable map can also be applied to the reference field degrees of freedom, for $T^\mu$ we get ${\mathcal O}_{f,T^\mu}(\tau^\mu)=\tau^\mu$ which demonstrates that the reference fields are no dynamical degrees of freedom in the reduced phase space. The independent elementary variables of the reduced phase space are just given by the observables of all but the clock degrees of freedom excluding also lapse and shift degrees of freedom. The Poisson algebra of the observables can be shown to be given by
\begin{equation}
\label{obsAlgegra}
\{{\mathcal O}_{f,T^\mu}, {\mathcal O}_{g,T^\mu}\}\simeq {\mathcal O}_{\{f,g\}^*,T^\mu},   
\end{equation}
where $\{f,g\}^*$ denotes the Dirac bracket associated with the second class system of the constraints $({\mathcal G}^\mu,\tilde{c}_\mu)$. If we apply the observable map in (\ref{obsmap}) on lapse and shift variables these are automatically mapped to functions on the reduced phase space, whereas their momenta are mapped to the primary constraints, which we will discuss for the dust models under consideration in subsections \ref{sec:RelDustGD} and \ref{sec:RelDustBK}. 

Besides the kinematics that is encoded in the observable algebra in (\ref{obsAlgegra}), of further interest is the dynamics of the observables. Since by construction they commute with all constraints the dynamics can obviously not be generated by $H_{\rm ext}$. Instead their dynamics is described as an evolution with respect to physical time $\tau^0$ and generated by a so called physical Hamiltonian ${\bf H}_{\rm phys}$. In this work we will only consider reference fields that lead to a deparamatrization of the Hamiltonian constraint. In this case, the Hamiltonian constraint can be written linearly in the temporal reference field momentum $P_0$, that is 
\bq
\lb{deparametrization}
\tilde c =P_0+h(q_{ab}, p^{ab},\varphi,\pi_\varphi,T^0_{,a}),
\eq
where the function $h$ can depend on all phase space variables but the reference field momenta and the spatial reference fields $T^a$ and if it depends on the the temporal reference field $T^0$, then only via its spatial derivatives. That this is satisfied for the dust reference fields considered here will be shown below. Then the physical time evolution of $\mathcal O_{f, T^\mu}(\tau^\mu)$ can be easily formulated as 
\bq
\lb{eomObs}
\frac{d \mathcal O_{f, T^\mu}}{d\tau^0}(\tau^\mu)=\Big\{ \mathcal O_{f, T^\mu}(\tau^\mu),  {\bf H}_{\rm phys}\Big\},\quad {\rm with}\quad
{\bf H}_{\rm phys}:=\int\limits_{\cal S}d^3\tau  \mathcal{O}_{h, T^\mu}(\vec{\tau}),
\eq
where ${\cal S}$ denotes the spatial reference field manifold with coordinates $\tau^j$, j=1,2,3. 

A crucial property of the class of deperamatrized models is that the physical Hamiltonian does not depend on the physical time $\tau^0$, which is ensured by the restrictions we made for the function $h$ in (\ref{deparametrization}). As shown in \cite{Vy1994,dittrich,thiemann2006} the multiparameter family of maps  ${\cal O}^{\tau^\mu}: f \mapsto {\cal O}_{f,T ^\mu}(\tau^\mu)$ is a homomorphism from the commutative algebra of functions on phase space to the commutative algebra of weak Dirac observables, both with pointwise multiplication, that is
\begin{equation*}
{\cal O}_{f,T^\mu}(\tau^\mu)+ {\cal O}_{g,T^\mu}(\tau^\mu) \simeq 
{\cal O}_{f+g,T^\mu}(\tau^\mu)\quad {\rm and}\quad
{\cal O}_{f,T^\mu}(\tau^\mu){\cal O}_{g,T^\mu}(\tau^\mu) \simeq 
{\cal O}_{fg,T^\mu}(\tau^\mu) ~.\
\end{equation*}
This can be used to easily obtain the physical Hamiltonian ${\bf H}_{\rm phys}$ once the observables for the elementary phase space variables have been constructed using
\begin{equation*}
 {\bf H}_{\rm phys}=  \int\limits_{\cal S}d^3\tau H(\vec{\tau}),\quad
 H:={\cal O}_{h,T^\mu}=h(O_{q_{ab},T^\mu},O_{p^{ab},T^\mu},O_{\varphi,T^\mu},O_{\pi_\varphi,T^\mu},O_{T_{,a},T^\mu}) ~.
\end{equation*}
This finishes the brief review on the relational formalism and the construction of observables for generic reference fields. In the next two subsections we will present two dust models that will be used as reference fields in the remaining part of the article, these are the Gaussian and the Brown-Kucha\v{r} dust models.

\subsection{The relational formalism with Gaussian dust}
\label{sec:RelDustGD}
The Gaussian dust was first proposed and discussed in \cite{kt1991}. The action of the Gaussian dust is given by 
\bq
S_\mathrm{G}=-\int d^3x \sqrt{-g}\left(\frac{\rho}{2}[g^{\mu\nu}T_{,\mu}T_{,\nu}+1]+g^{\mu\nu}T_{,\mu}W_jS^j_\nu\right),
\eq
which involves eight dynamical dust fields $T, S^j,\rho,W_j$ with $j=1,2,3$. The equations of motion of $W_j,\rho$ impose the following conditions on the metric 
\bq
g^{\mu\nu}T_{,\mu}T_{,\nu}+1=0, \quad \quad g^{\mu\nu}T_{,\mu}S^j_\nu=0.
\eq
The system $S_{\rm tot}=S_{\rm geo}+S_{\rm scalar}+S_{\rm G}$ is no longer first class but a second class system. In the extended phase space, besides the 22 degrees of freedom coming from the geometrical and matter sectors, there are 16 additional degrees of freedom due to the dust fields which are $T$, $\rho$, $S^i$, $W_j$ and their respective conjugate momenta. As shown in  \cite{gt2015} there are 8 second class constraints which can be used to  remove the dynamical fields $\rho$ and $W_j$ and their momenta completely from the  second class reduced phase space. So the remaining canonical variables from the dust fields are $T^\mu=(T, S^j)$ and their respective conjugate momenta $P,P_j$ that will be used as reference fields for the Hamiltonian and spatial diffeomorphism constraints respectively.  Finally, the Hamiltonian and diffeomorphism constraints  of the coupled system take the form
\bqn
c^\mathrm{tot}&=&c+c^\mathrm{dust},\\
c^\mathrm{tot}_a&=&c_a+c^\mathrm{dust}_a,
\eqn 
where $c=c^\mathrm{geo}+c^\mathrm{scalar}$, $c_a=c^\mathrm{geo}_a+c^\mathrm{scalar}_a$ are given in  (\ref{2a2})-(\ref{2a5}) and the dust contributions are given by 
\bqn
\lb{gdust1}
c^\mathrm{dust}&=&P\sqrt{1+q^{ab}T_{,a}T_{,b}}+\frac{q^{ab}T_{,a}P_jS^j_{,b}}{\sqrt{1+q^{ab}T_{,a}T_{,b}}},\\
\lb{gdust2}
c^\mathrm{dust}_a&=&PT_{,a}+P_jS^j_{,a} .
\eqn
In order to satisfy the condition that the reference fields $(T,S^j)$ are at least weakly canonically conjugate to the constraints $c^{\rm tot},c_a^{\rm tot}$ we solve the constraints $c^\mathrm{tot}$ and $c^\mathrm{tot}_a$  for the dust momenta $(P,P_j)$ yielding the following equivalent set of constraints 
\bqn
\lb{de1}
\tilde c^\mathrm{tot}&=&P+c\sqrt{1+q^{ab}T_{,a}T_{,b}}-q^{ab}T_{,a}c_b=:P+h^{\rm G},\\
\lb{de2}
\tilde c^\mathrm{tot}_j&=&P_j+S^a_j\left(-h^{\rm G} T_{,a}+c_a\right),
\eqn 
where $S^a_j$ is the inverse of $S^j_{,a}$ with $S_{,a}^jS^a_k=\delta^j_k$ and $S^j_{,a}S^b_j=\delta_a^b$ and $h^{\rm G}$ is defined by $h^{\rm G}\coloneqq c\sqrt{1+q^{ab}T_{,a}T_{,b}}-q^{ab}T_{,a}c_b$, where the superscript `$\rm G$' stands for Gaussian dust. 

Since the constraints $\tilde{c}_I=(\tilde{c}^{\rm tot}_\mu,\pi_\mu)$ mutually Poisson commute we can separate the construction of the observables in two steps: first with respect to spatial diffeomorphisms $\tilde{c}_j$ and then with respect to the remaining constraints. The symplectic reduction of the spatial diffeomorphism constraints is implemented by pulling back all tensors by the diffeomorphism $x\mapsto \sigma^j=S_j
(x)$, which as shown in \cite{bk1995,ghtw2010I} is a canonical transformation for $(q_{ab},p^{ab},N,\pi,{N}^a,\pi_a,\varphi,\pi_\varphi,T,P)$. The canonical conjugate momenta to $S^j$ become $\tilde{c}^{\rm tot}_j=S^a_jc_a^{\rm tot}$ and thus this canonical pair drops out of the partially reduced phase space. Let us denote the transformed quantities by $({q}^\prime _{ij},{p^\prime}^{ij},N',\pi^\prime,{N^\prime}^j,\pi^\prime_j,\varphi^\prime,\tilde{\pi}^\prime_\varphi,T^\prime,P^\prime)$ in order to distinguish them from the original variables. Here the primed variables are all invariant under spatial diffeomorphisms. Note that due to the pullback these are scalars on the original manifold $\sigma$ but tensors on the dust manifold that we denote by ${\cal S}$. The indices $i,j,k\dots=1,2,3$ refer to tensor indices on ${\cal S}$. Furthermore, we also have to pullback $\tilde{c}^{\rm tot}$ leading to $\tilde{c}^{\prime\rm tot}=\tilde{c}^{\rm tot}({q}^\prime _{ij},{p^\prime}^{ij},\varphi^\prime,\tilde{\pi}^\prime_\varphi,T^\prime_{,j},P^\prime)$.
To obtain the final observables we apply the observable map in (\ref{obsmap}) for the remaining constraints with the gauge fixing conditions ${\cal G}^0=\tau-T, {\cal G}^j=\sigma^j-S^j$, where the latter is only relevant for $\dot{\cal G}^j$ being involved in the symplectic reductions with respect to $\pi^\prime_j$. In order to keep our notation simple we denote the final observables just by capital letters and get 
\bq
\lb{ob}
Q_{ij}\coloneqq \mathcal O_{q^\prime_{jk}, T}(\tau, \vec \sigma), \quad P^{ij}\coloneqq \mathcal O_{{p^\prime}^{jk}, T}(\tau, \vec \sigma), \quad \Phi \coloneqq \mathcal O_{\varphi^\prime, T}(\tau, \vec \sigma), \quad \Pi_\Phi \coloneqq \mathcal O_{\pi^\prime_\varphi, T^\mu}(\tau, \vec \sigma). 
\eq
If we apply the  observable map in (\ref{obsmap}) also to lapse and shift degrees of freedom one can easily find that 
\bq
\mathcal O_{N^\prime, T}(\tau, \vec \sigma)=1,\quad \mathcal O_{{N^\prime}^j, T}(\tau, \vec \sigma)=0, \quad \mathcal O_{\pi^\prime, T}(\tau, \vec \sigma)=\pi^\prime,\quad 
\mathcal O_{\pi^\prime_j, T}(\tau, \vec \sigma)=\pi^\prime_j.
\eq
As discussed above the physical Hamiltonian ${\bf H}^{\rm G}_{\rm phys}$ for the Gaussian dust model is given by
\bq
\lb{gaussham}
{\bf H}^{\rm G}_\mathrm{phys}=\frac{1}{\kappa}\int\limits_{\cal S}d^3\sigma\,
{\cal O}_{{h^\prime}^{\rm G},T}=\frac{1}{\kappa}\int\limits_{\cal S} d^3\sigma\, C,\quad C:={\cal O}_{c^\prime,T}
\eq
with $c^\prime$ being the spatially diffeomorphism invariant version of  $c=c^\mathrm{geo}+c^\mathrm{scalar}$ given in (\ref{2a2})-(\ref{2a3}). To obtain $C$ we only need to replace the dynamical variables in those equations with their respective observables  defined in (\ref{ob}). More specifically, 
\bq
C=\frac{1}{\sqrt{Q}}G_{ijmn}P^{ij}P^{mn}-\sqrt{Q}R^{(3)}+\frac{\kappa}{2}\Bigg\{\frac{\lambda_\varphi\Pi^2_\Phi}{\sqrt{Q}}+\frac{\sqrt{Q}}{\lambda_\varphi}\left(Q^{ij}\partial_i \Phi \partial_j \Phi+V\right)\Bigg\},
\eq
where 
\bq
G_{ijmn}=\frac{1}{2}\left(Q_{im}Q_{jn}+Q_{in}Q_{jm}-Q_{ij}Q_{mn}\right).
\eq
The elementary Poisson brackets of the canonical variables on the reduced phase space can be obtained from (\ref{obsAlgegra}). The Gaussian dust model has the property that all original variables but the dust degrees of freedom Poisson commute with all gauge fixing constraints. As a consequence, for these variables the Dirac bracket in (\ref{obsAlgegra}) reduces to the usual Poisson bracket and this allows to easily compute the observable algebra. The only non-vanishing Poisson brackets are given by
\bqn
\lb{poisson1}
\{Q_{ij}(\tau,\vec \sigma),P^{kl}(\tau,\vec \sigma')\}&=&\kappa \delta^k_{(i}\delta^l_{j)}\delta^3(\vec \sigma-\vec \sigma'),\\
\lb{poisson2}
\{\Phi(\tau,\vec \sigma), \Pi_\Phi(\tau,\vec \sigma')\}&=&\delta^3(\vec \sigma-\vec \sigma') .
\eqn

Given this physical Hamiltonian we can also compute the physical time evolution of a generic function on the reduced phase space 
\bq
\frac{dF}{d\tau}(Q_{ij},P^{ij},\Phi, \Pi_\Phi)=\{F, {\bf H}^{\rm G}_\mathrm{phys}\},
\eq
yielding for the elementary canonical observables in the Gaussian dust model the following equations of motion:
\bqn
\lb{gausseom1}
\dot \Phi &=& \frac{\lambda_\varphi  \Pi_\Phi}{\sqrt{Q}}, \\
\dot \Pi_\Phi&=&-\frac{\sqrt{Q}V_{,\Phi}}{2\lambda_\varphi}+\partial_j\left(\frac{\sqrt{Q}Q^{ij}}{\lambda_\varphi}\partial_i \Phi\right),\\
\dot Q_{ij}&=&\frac{2}{\sqrt{Q}}G_{ijmn}P^{mn},\\
\lb{gausseom2}
\dot P^{ij}&=&\frac{Q^{ij}}{2\sqrt Q}G_{klmn}P^{kl}P^{mn}+\frac{1}{\sqrt Q}\left(\mathrm{tr}PP^{ij}-2 P^{ik}P^j_k\right)+\frac{1}{2}\sqrt{Q}Q^{ij}R^{(3)}-\sqrt{Q}R^{ij}\nb\\
&&-\frac{\kappa \sqrt Q }{4\lambda_\varphi}\left(Q^{ij}D^k\Phi D_k\Phi -2D^i\Phi D^j\Phi +Q^{ij} V\right)+\frac{\kappa \lambda_\varphi Q^{ij}}{4\sqrt Q}\Pi^2_\Phi.
\eqn
In the next section we will discuss a second so called Brown-Kucha\v{r} dust model that will also be used as reference fields in the remaining part of the paper.

\subsection{The relational formalism with  Brown-Kucha\v{r} dust}
\label{sec:RelDustBK}
The Brown-Kucha\v{r} dust  was first proposed and discussed in \cite{bk1995}. The action of the Brown-Kucha\v{r} dust is given by 
\bq
S_\mathrm{BK}=-\frac{1}{2}\int_\mathcal M d^4x \sqrt{-g} \, \rho \, [g^{\mu\nu}U_\mu U_\nu+1],
\eq
where $\rho$ is the energy density of the dust fields  and the unit time-like dust velocity field can be expressed in terms of the elementary fields as $U=-dT+W_j dS^j$. Similar to the Gaussian dust, the dynamical fields of the Brown-Kucha\v{r} dust consist of $T$, $S^j$, $\rho$,$W_j$  and their respective conjugate momenta $P$, $P_j$, $\pi_\rho$ and $\pi_{W_j}$.  A detailed analysis of the Hamiltonian structure of the whole system in \cite{ghtw2010I} yields altogether 8 first-class constraints and 8 second-class constraints,  leaving after the reduction of the 8 second class constraints 30 dynamical degrees of freedom in the phase space, which are the same elementary variables as in the Gaussian dust model. As before,  $\rho$, $W_j$ and their respective momenta can be expressed in terms of the remaining variables and the system becomes first class with the following form of  Hamiltonian and diffeomorphism constraints  \cite{bk1995,thiemann2006}
\bqn
\lb{bkdust1}
c^\mathrm{tot}&=&c+c^\mathrm{dust}=c+P \sqrt{1+q^{ab}U_aU_b},\\
\lb{bkdust2}
c^\mathrm{tot}_a&=&c_a+c^\mathrm{dust}_a=c_a+ PT_{,a}+P_j S^j_{,a},
\eqn
where $c$ and $c_a$ are given in  (\ref{2a2})-(\ref{2a5}), $U_a:=T_{,a}+W_jS^j_{,a}$ and the contribution of the dust to the total constraint was taking into account that has been derived in \cite{bk1995,thiemann2006} . 

Compared to the Gaussian dust contribution in  (\ref{gdust1})-(\ref{gdust2}), the difference between two dust fields lies in the specific form of $c^\mathrm{dust}$. More substantial differences can be seen  when the Hamiltonian constraint $c^\mathrm{tot}$ is rewritten in deparametrized form. As in the case of the Gaussian dust model we can solve $c^\mathrm{tot}$  for $P$ and $c^\mathrm{tot}_a$ for $P_j$ leading to 
 \bqn
 \lb{2c1}
 \tilde c^\mathrm{tot}&=&P-\mathrm{sgn}(P)\sqrt{c^2-q^{ab}c_ac_b}=:P-\mathrm{sgn}(P)h^{\rm BK}, \\
 \lb{2c2}
 \tilde{c}^{\rm tot}_j&=&P_j+S^a_j\left(-h^{\rm G} T_{,a}+c_a\right),
 \eqn
where $\mathrm{sgn} (P)$ denotes the sign of $P$ and $h^{\rm BK}:=\sqrt{c^2-q^{ab}c_ac_b}$. Here the superscript `$\rm BK$' is introduced to denote the Brown-Kucha\v{r} model. For the Hamiltonian constraint, the Brown-Kucha\v{r} mechanism has been used in order to write it in deparametrized form, see \cite{ghtw2010I}. Similar to the Gaussian dust model the spatial diffeomorphism constraint does not deparametrize but this does not make the construction of the observables more difficult. The latter can be constructed in the same way as it was already discussed in Sec. \ref{sec:RelDustGD} for the Gaussian dust model (as in Gaussian dust case, we denote the final observables by capital letters). The only difference to the former model is that ${h^\prime}^{GD}$ needs to be replaced by ${h^\prime}^{BK}$ in the observable map (\ref{obsmap}) that involve $\tilde{c}^{\prime{\rm tot}}$. The details of the construction can be found in \cite{ghtw2010I}. Following the same line of argument as in the Gaussian  dust case, the physical Hamiltonian  is the observable associated with ${h^\prime}^{\rm BK}$, which yields\footnote{Here, we choose $\mathrm{sgn}(P)$ to be negative so that the Hamiltonian (\ref{bkham}) is bounded from below.}
\bq
\lb{bkham}
{\bf H}^{\rm BK}_\mathrm{phys}=\frac{1}{\kappa}\int\limits_{\cal S} d^3\sigma\,
O_{{h^\prime}^{\rm BK},T}
=\frac{1}{\kappa}\int\limits_{\cal S}\, d^3\sigma\,\sqrt{C^2-Q^{ij}C_iC_j}=:\frac{1}{\kappa}\int\limits_{\cal S}\, d^3\sigma\, H(\sigma) .
\eq
Here $C=C^{\mathrm{geo}}+C^{\mathrm{scalar}}$, $C_i=C_i^{\mathrm{geo}}+C_i^{\mathrm{scalar}}$, where $H:=\sqrt{C^2-Q^{ij}C_iC_j}$ denotes the physical Hamiltonian density, and 
\bqn
C^{\mathrm{geo}}&=&\frac{1}{\sqrt{Q}}G_{ijmn}P^{ij}P^{mn}-\sqrt{Q}R^{(3)}\\
C^{\mathrm{scalar}}&=&\frac{\kappa}{2}\Bigg\{\frac{\lambda_\varphi\Pi^2_\Phi}{\sqrt{Q}}+\frac{\sqrt{Q}}{\lambda_\varphi}\left(Q^{ij}\partial_i \Phi \partial_j \Phi+V\right)\Bigg\}\\
C_i^{\mathrm{geo}}&=&-2 D_k P^k_i, \quad \quad C_i^{\mathrm{scalar}}=\kappa\Pi_\Phi \partial_i \Phi,
\eqn
with 
\bq
G_{ijmn}=\frac{1}{2}\left(Q_{im}Q_{jn}+Q_{in}Q_{jm}-Q_{ij}Q_{mn}\right).
\eq
The application of the observable map in (\ref{obsmap}) to lapse and shift degrees of freedom leads for the Brown-Kucha\v{r} model to the following observables:
\bq
\label{eq:ObsNNa}
\mathcal O_{N^\prime, T}(\tau, \vec \sigma)=\frac{C}{H}=\sqrt{1+\frac{Q^{ij}C_iC_j}{H^2}},\quad \mathcal O_{{N^\prime}^j, T}(\tau, \vec \sigma)=0, \quad \mathcal O_{\pi^\prime, T}(\tau, \vec \sigma)=\pi^\prime,\quad 
\mathcal O_{\pi^\prime_j, T}(\tau, \vec \sigma)=\pi^\prime_j.
\eq
As expected for the Brown-Kucha\v{r} model, the lapse and shift degrees of freedom can be completely expressed in terms of the physical variables and the primary constraints\footnote{In \cite{ghtw2010I,ghtw2010II} a shift vector in terms of physical degrees of freedom is defined as $N^j=-\frac{Q^{jk}C_k}{H}$ whereas in (\ref{eq:ObsNNa}) we present the Dirac observable associated to the shift vector which is just zero. Note that their is no contradiction in the two results because the $N
^j$ from \cite{ghtw2010I,ghtw2010II} is not the Dirac observable of the shift vector but has been defined as  follows: Consider $\{F,H(\sigma)\}=\frac{C}{H}\{F,C\}-\frac{Q^{jk}C_k}{H}\{F,C_j\}$. Then $N^j$ was defined as being the coefficient in front of the Poisson bracket of $\{F,C_j\}$.}
On the reduced phase space that is spanned by the physical degrees of freedom the only non-vanishing Poisson brackets are 
\bqn
\{Q_{ij}(\tau,\vec \sigma),P^{kl}(\tau,\vec \sigma^\prime\}&=&\kappa \delta^k_{(i}\delta^l_{j)}\delta^3(\vec \sigma-\vec \sigma^\prime),\\
\{\Phi(\tau,\vec \sigma), \Pi_\Phi(\tau,\vec \sigma^\prime)\}&=&\delta^3(\vec \sigma-\vec \sigma^\prime).
\eqn
As a result, the evolution of a generic phase space function $F(Q_{ij}, P^{ij}, \Phi, \Pi_\Phi)$ are governed by Hamilton's equation generated by ${\bf H}_{\rm phys}^{\rm BK}$
\bq
\lb{ham}
\frac{dF}{d\tau}(Q_{ij}, P^{ij}, \Phi, \Pi_\Phi)
=\{F, {\bf H}_\mathrm{phys}\} ~.
\eq
Hamilton's equations of the canonical variables can thus be computed in a straightforward way, here we simply list the results as follows \cite{ghtw2010I}:
\bqn
\lb{bkeom1}
\dot \Phi &=& \frac{\lambda_\varphi N \Pi_\Phi}{\sqrt{Q}}+N^i\partial_i \Phi, \\
\dot \Pi_\Phi&=&-\frac{N\sqrt{Q}V_{,\Phi}}{2\lambda_\varphi}+\partial_j\left(\frac{N\sqrt{Q}Q^{ij}}{\lambda_\varphi}\partial_i \Phi+\Pi_\Phi N^j\right),\\
\dot Q_{ij}&=&\frac{2N}{\sqrt{Q}}G_{ijmn}P^{mn}+2\partial_{(i}N_{j)}-2\Gamma^k_{ij}N_k,\\
\lb{bkeom2}
\dot P^{ij}&=&N\Bigg\{\frac{Q^{ij}}{2\sqrt Q}G_{klmn}P^{kl}P^{mn}+\frac{1}{\sqrt Q}\left(Q\cdot P P^{ij}-2 P^{ik}P^j_k\right)+\frac{1}{2}\sqrt{Q}Q^{ij}R^{(3)}-\sqrt{Q}R^{ij}\Bigg\}\nb\\
&&-\sqrt{Q}Q^{ij}D^kD_kN+\sqrt Q D^i D^j N+D_k\left(N^kP^{ij}\right)-D_k N^i P^{jk}-D_k N^j P^{ik}\nb\\ 
&&-\frac{\kappa \sqrt Q N}{4\lambda_\varphi}\left(Q^{ij}D^k\Phi D_k\Phi -2D^i\Phi D^j\Phi +Q^{ij} V\right)+\frac{N\kappa \lambda_\varphi Q^{ij}}{4\sqrt Q}\Pi^2_\Phi-\frac{\kappa}{2}\mathcal HN^iN^j.
\eqn
Let us briefly comment on the negative dust energy density that was originally chosen for the dust reference fields in \cite{ghtw2010II}. Whether one chooses a positive or negative dust energy density crucially depends on the part of phase space that one wants to include in the deparameterized model. For instance, if Minkowski flat space should be part of the phase space with the standard model matter, then the physical Hamiltonian needs to be positive and consequently the energy density of the dust negative in order that the total Hamiltonian constraint can be satisfied, see \cite{ghtw2010I} for more details. 
However, in the case of cosmology the gravitational contribution is negative and therefore both options negative and positive dust energy density are consistent with the constraint, hence in our work here we will discuss both cases.

 Summarizing this section, we have given a review on Hamilton's equations in the relational formalism when Gaussian dust and Brown-Kucha\v{r} dust serve as the reference fields. Both Gaussian and Brown-Kucha\v{r} dust fields introduce four additional physical degrees of freedom to the original system, which is composed of 3 physical degrees of freedom in the configuration space.  The most important results in this section are Hamilton's equations (\ref{gausseom1})-(\ref{gausseom2}) in the Gaussian dust model and Hamilton's equations  (\ref{bkeom1})-(\ref{bkeom2}) in the Brown-Kucha\v{r} dust model. Comparing the two sets of equations, one can find that Hamilton's equations in the Brown-Kucha\v{r} dust model can reduce to those in the Gaussian dust model when the lapse is set to unity and shift vanishes. As a result, if we consider a spatially-flat FLRW universe, the two sets  of equations predict the same background dynamics, the difference will only occur at the level of the perturbations. 

\section{The background dynamics in a spatially-flat FLRW universe with dust reference fields}
\label{sec:BackGround}
\renewcommand{\theequation}{3.\arabic{equation}}\setcounter{equation}{0}

In this section, based on Hamilton's equations given in the last section, we  consider the background dynamics of a spatially-flat FLRW universe with a Klein-Gordon scalar field in the relational formalism using a dust reference field. Due to the homogeneity and isotropy, the image of the spatial background metric under the observable map carries only one physical degree of freedom which can be parametrized by $\overline Q_{ij}=A(\tau)\delta_{ij}$, where $A(\tau)$ is the image of the squared scale factor under the observable map, namely, $A(\tau) ={\mathcal O}_{a^2,T^\mu}(\tau)$.   The image of the conjugate momentum of the background ADM-metric under the observable map can be written as  
$\overline P^{ij}=P(\tau)\delta^{ij}$. In addition there is another physical degree of freedom encoded in $\overline{\Phi}$ with its canonically conjugate momentum $\overline{\Pi}_\Phi$. To avoid confusion with the later used perturbed quantities, some of the background quantities  are denoted with a bar on the top. In terms of $A(\tau),P(\tau)$ and $\overline{\Phi}(\tau),\overline{\Pi}_\Phi(\tau)$,  the physical Hamiltonians  in  (\ref{gaussham}) and (\ref{bkham}) can  now be reduced to 
\bq
\lb{3a2}
\overline{\bf H}_\mathrm{phys}=\mathcal V_0 \overline C=\mathcal V_0\left(-\frac{3 P^2\sqrt{A}}{2\kappa}+\frac{\lambda_\varphi \overline  \Pi^2_\Phi}{2A^{3/2}}+\frac{A^{3/2}V(\overline{\Phi})}{2\lambda_\varphi}\right),
\eq
where $\mathcal V_0$ is the volume of the fiducial cell chosen to define symplectic structure or the finiteness of spatial integral in   (\ref{ham}).  The background phase space variables satisfy the following Poisson brackets 
\bqn
\lb{3a3}
\{A(t), P(t)\}=\frac{\kappa}{3\mathcal V_0}, ~~~~~
\{\overline \Phi(t), \overline \Pi_\Phi(t)\}=\frac{1}{\mathcal V_0} 
\eqn
with all remaining Poisson brackets vanishing. 

It is straightforward to find the equations of motion of the background variables by evaluating the Poisson bracket between dynamical variables and the physical Hamiltonian. This results in the following Hamilton's equations:
\bqn
\lb{3a4}
\dot A&=&- P\sqrt{A},\quad \dot {P}=\frac{P^2}{4\sqrt{A}}+\frac{\kappa \lambda_\varphi \overline{\Pi}^2_\Phi}{4A^{5/2}}-\frac{\kappa \sqrt{A}V}{4\lambda_\varphi},\\
\lb{3a7}
\dot{ \overline{\Phi}}&=&\frac{\lambda_\varphi \overline \Pi_\Phi}{A^{3/2}}, \quad \dot {\overline \Pi}_\Phi=-\frac{A^{3/2}V_{,\overline\Phi}}{2\lambda_\varphi}.
\eqn
With the  above Hamilton's equations (\ref{3a4})-(\ref{3a7}), one can easily find the Klein-Gordon and Raychaudhuri equations for $\overline \Phi$ and $A$, which are  given by 
\bqn
\lb{3a8}
\ddot {\overline \Phi}+3H\dot {\overline \Phi}&+&\frac{1}{2}V_{,\overline \Phi}=0,\\
\lb{3a9}
\frac{\ddot A}{2 A} - H^2  &=& - \frac{4\pi G}{3}\left(\rho+3p\right),
\eqn
where $H=\frac{\dot A}{2 A}$ is the Hubble rate and $\rho$ and $p$ denote the total energy density and pressure. These are  given respectively, by 
\bq
\lb{3a11}
\rho=\rho_\Phi+\rho_\mathrm{dust},\quad \quad p=p_\Phi+p_\mathrm{dust},
\eq
with 
\bqn
\lb{3a12}
\rho_\Phi&=&\frac{1}{2\lambda_\varphi}\left(\dot{\overline  \Phi}^2+V\right),\quad \quad
p_\Phi=\frac{1}{2\lambda_\varphi}\left(\dot{\overline  \Phi}^2-V\right),\nb\\
\rho_\mathrm{dust}&=&\frac{\overline {\mathcal E}^\mathrm{dust}}{A^{3/2}}=\frac{-\overline C}{A^{3/2}},\quad\quad  p_\mathrm{dust}=0 .
\eqn
Here we have used $\overline{C}^{\rm tot}=0=\overline{\mathcal E}^\mathrm{dust}+\overline{C}$. Note that $\overline{\mathcal E}^\mathrm{dust}$  is a constant of motion as can be shown using Hamilton's equations (\ref{3a4})-(\ref{3a7}). We have 
\begin{equation}
\overline {\mathcal E}^\mathrm{dust}=\frac{3 P^2\sqrt A}{2\kappa}-A^{3/2}\overline \rho_\Phi= -\overline C.
\end{equation}
The gauge-invariant Friedmann equation can be derived from $\overline C=-\overline{\mathcal E}^\mathrm{dust}$ by rearranging the terms and using the definitions in (\ref{3a12}), the result is 
\bq
\lb{Friedmann}
H^2=\frac{8\pi G}{3}\left(\rho_\Phi+\rho_\mathrm{dust}\right).
\eq
Finally, from equation (\ref{3a8}), one can find the continuity equation for  the scalar field, that is,
\bq
\dot \rho_\Phi+3H \left(\rho_\Phi+P_\Phi\right)=0.
\eq
Similarly, from equations  (\ref{3a9}) and (\ref{Friedmann}), it can be shown that the continuity equation also holds individually for the dust field.

Although these equations have the same forms as their counterparts in conventional cosmology without dust reference fields, they are actually expressed in terms of the physical observables in the relational formalism. These physical observables are directly constructed from  the observable map of the corresponding gauge-variant quantities. We want to emphasize that the  similarity is only restricted to the forms of the equations. When implementing these equations, there are substantial differences in the way the physical Hamiltonian in (\ref{3a2}) is treated. In particular we have a true Hamiltonian here that is not vanishing on the constraint surface and both $A(\tau)$ and $\overline \Phi(\tau)$ as well as their corresponding momenta are independent physical degrees of freedom. Without the dust reference field one ends up with one independent physical degree of freedom once the symplectic reduction with respect to $\overline{C}^{\rm tot}$ has been performed.

We now proceed with  the numerical solutions of the background dynamics by integrating the Klein-Gordon equation, the Friedmann equation and the continuity equation of the dust field. In the numerics, we use Planck units and initial conditions are chosen at $\tau=0$. The parameter space of the solutions consists of four variables, namely, $(A_0, \overline \Phi_0, \dot{\overline \Phi}_0, \rho_{0,\mathrm{dust}})$. The self-interaction of the scalar field is taken to be the Starobinsky potential  
\bq
\lb{starobinsky}
V=\frac{3m^2}{16\pi G}\left(1-e^{-\sqrt{\frac{16\pi G}{3}}\overline \Phi}\right)^2,
\eq
where the mass is fixed to $2.49\times 10^{-6} $ from Planck results \cite{lsw2019}.  In the following, we will give three examples of our numerical solutions which mainly focus on the situations when the inflaton initially climbs up the inflationary potential with a positive velocity. The situation in which the inflaton rolls down the inflationary potential with a negative velocity from the start will be commented in the concluding remarks.

\begin{figure}
\includegraphics[width=8cm]{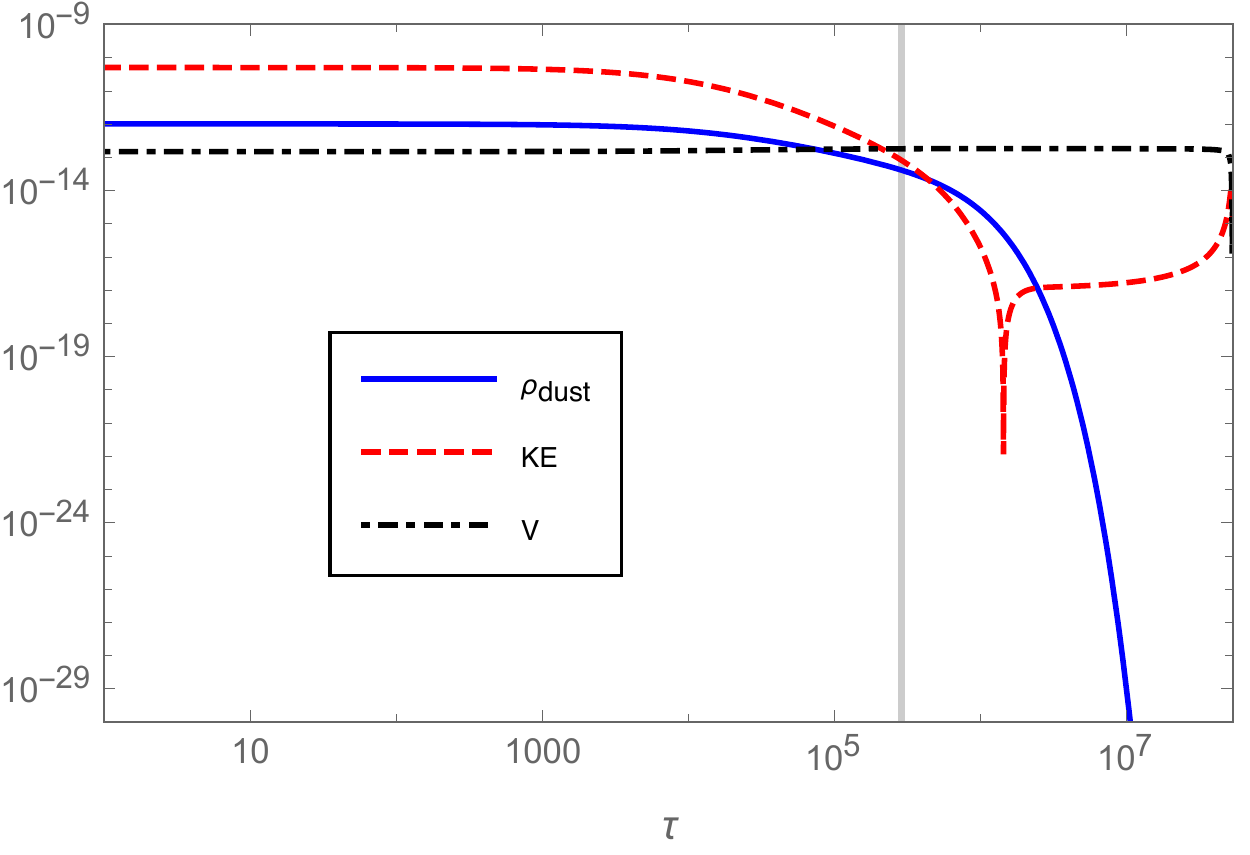}
\includegraphics[width=8cm]{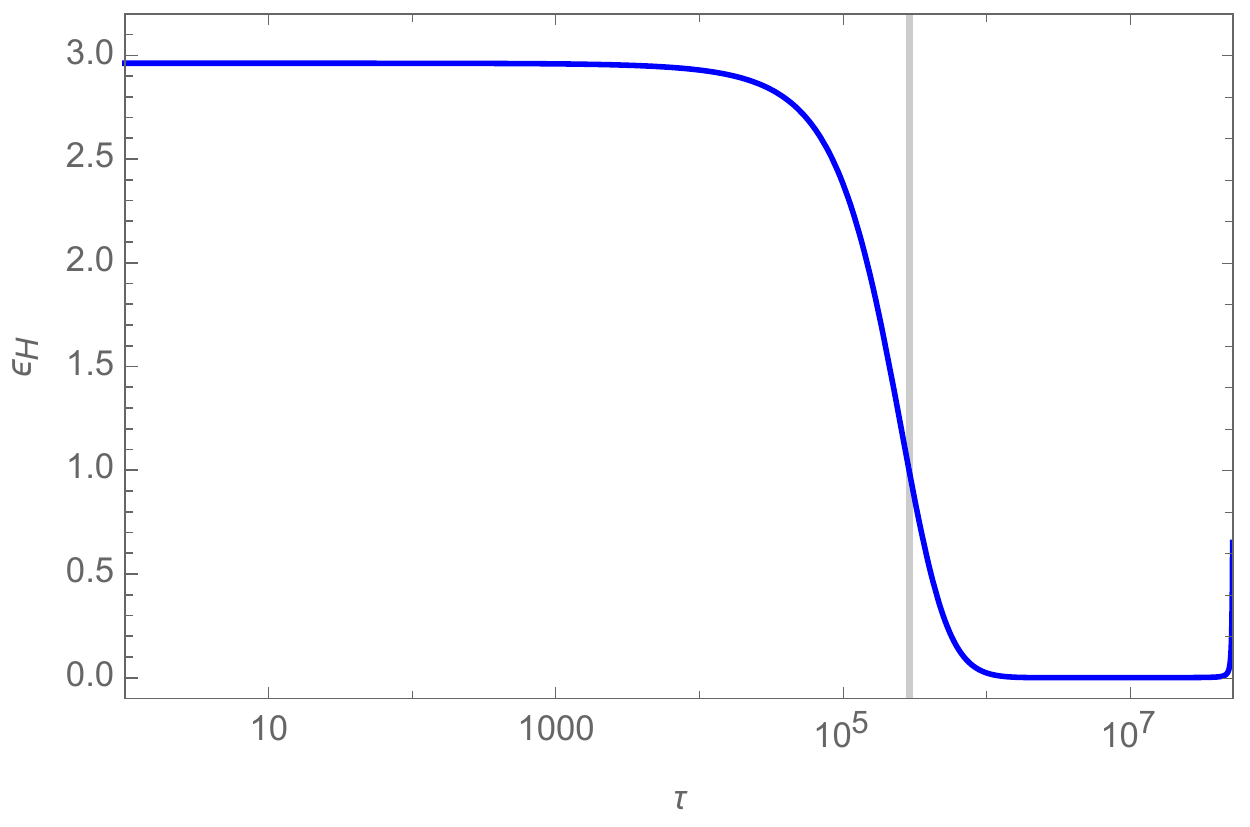}
{\caption{In this figure, with the initial conditions given in  (\ref{initial1}), evolution of the energy density of dust fields (blue line), the kinetic energy of the scalar field (red dashed line) and the potential energy (black dotdashed line) of the scalar field  is depicted in the left panel. The right panel shows the evolution of the first slow-roll parameter $\epsilon_H$ until the end of inflation. The vertical lines in these graphs indicate the onset of inflation at dust time $\tau=2.90\times10^5$.  The inflation  ends at  $\tau= 5.22\times10^7$,  yielding $61.6$ inflationary e-foldings.} 
\label{f1}}
\end{figure}

The first example is given in Fig. \ref{f1},  where the initial conditions are set in Planck units to  
\bqn
\lb{initial1}
A_0&=&1.50\times10^3, \quad \overline \Phi_0=0.540, \quad  \dot{\overline \Phi}_0=10^{-5}, \nb\\
\rho_{0,\Phi}&=&5.01\times10^{-11}, \quad \rho_{0,\mathrm{dust}}=10^{-12},\quad  \overline {\mathcal E}^{\mathrm{dust}}=5.81\times10^{-8}.
\eqn
Under these initial conditions, as shown in Fig. \ref{f1}, the universe is initially dominated by the kinetic energy of the scalar field (given by $\dot{\overline \Phi}^2/2$) and  inflation takes place when the first slow-roll parameter
\bq
\lb{4c1}
\epsilon_H=\frac{4\pi G}{H^2}\left(\rho_\Phi+p_\Phi+\rho_\mathrm{dust}\right),
\eq
becomes less than unity. The total inflationary e-foldings turns out to be 61.6. For these initial conditions, during the evolution of the universe, the contribution from the dust fields is always subdominant.  Furthermore, the energy density of the dust field decays rather quickly during slow-roll as it is inversely proportional to  the volume  of the universe.  
In order to study  the impact of the dust fields on the background dynamics, we also obtained numerical solutions with almost the same initial conditions  as in  (\ref{initial1}) but in absence of dust reference fields.  In this case, the inflationary  
e-foldings turns out to be approximately $68.0$ which shows the dust fields with positive energy density can diminish the number of inflationary e-folds. The reason for this is tied to an increase in the Hubble friction while the field climbs up the Starobinsky potential in presence of positive dust energy density resulting in a slightly smaller value of the inflaton field at the onset of inflation. 

\begin{figure}
{
\includegraphics[width=8cm]{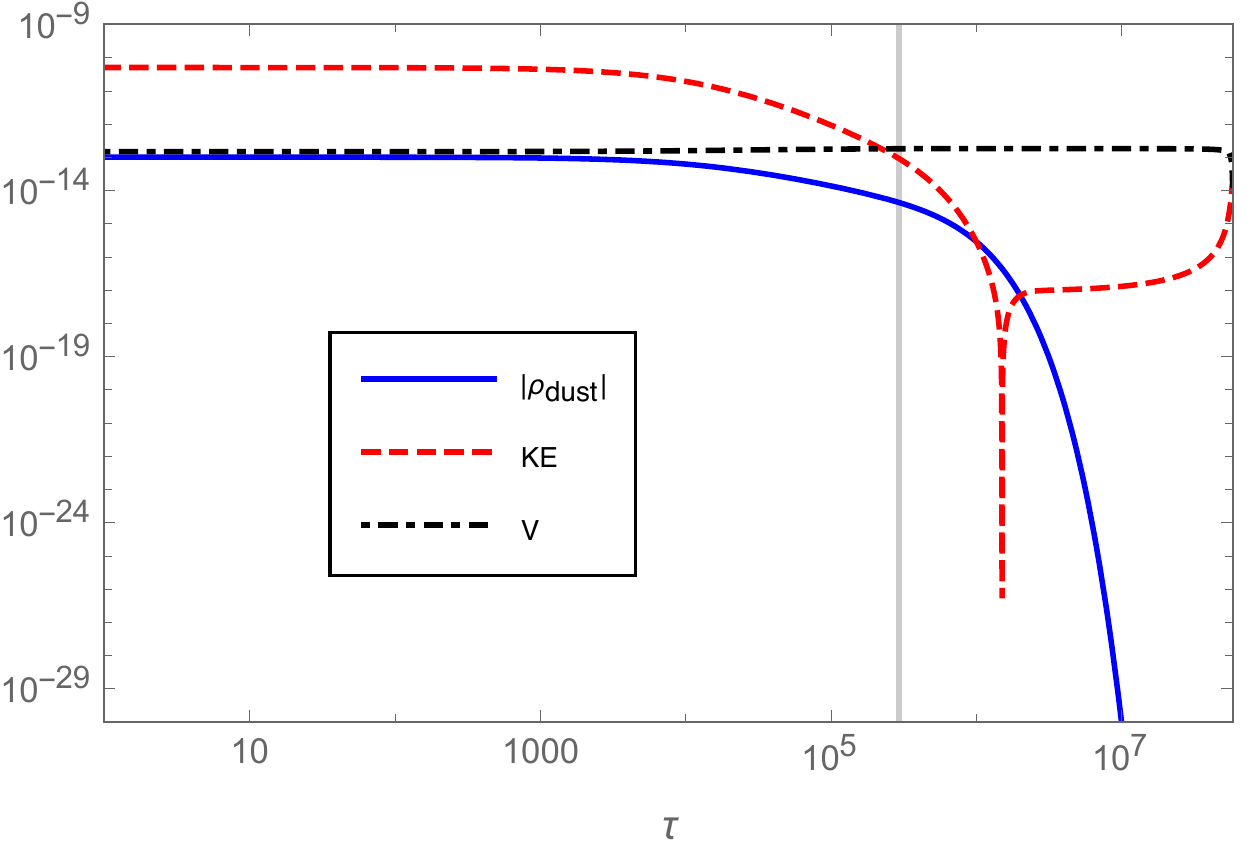}
\includegraphics[width=8cm]{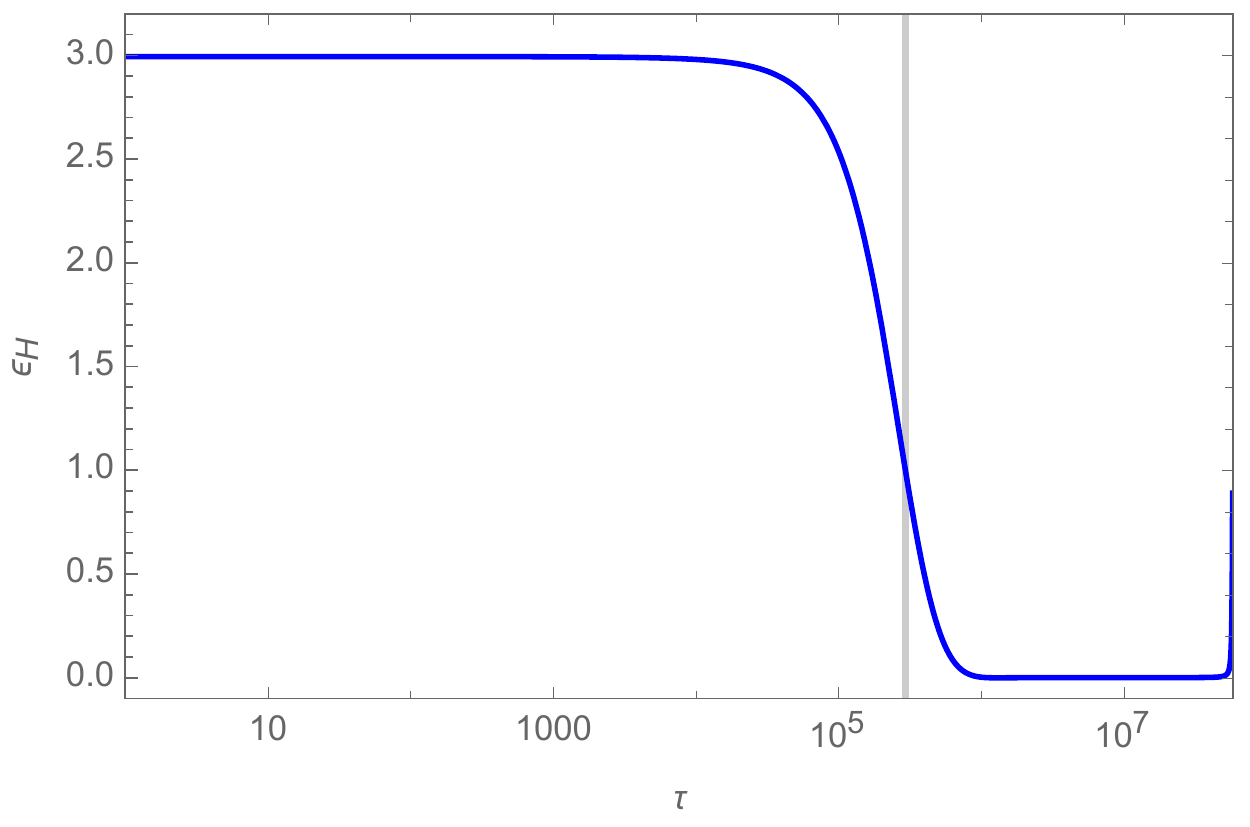}
}
\caption{With the initial conditions given by (\ref{initial3}) with a negative energy density for dust, the slow-roll phase can still take place if the universe is dominated by the scalar field at all times. The inflation occurs at $\tau=2.95\times10^5$ and ends at $\tau=5.81\times10^7$. In this case, the total number of the  inflationary e-folding is $68.8$. The vertical lines mark the onset of inflation in each subfigure.}
\label{f3}
\end{figure}

\begin{figure}
{
\includegraphics[width=8cm]{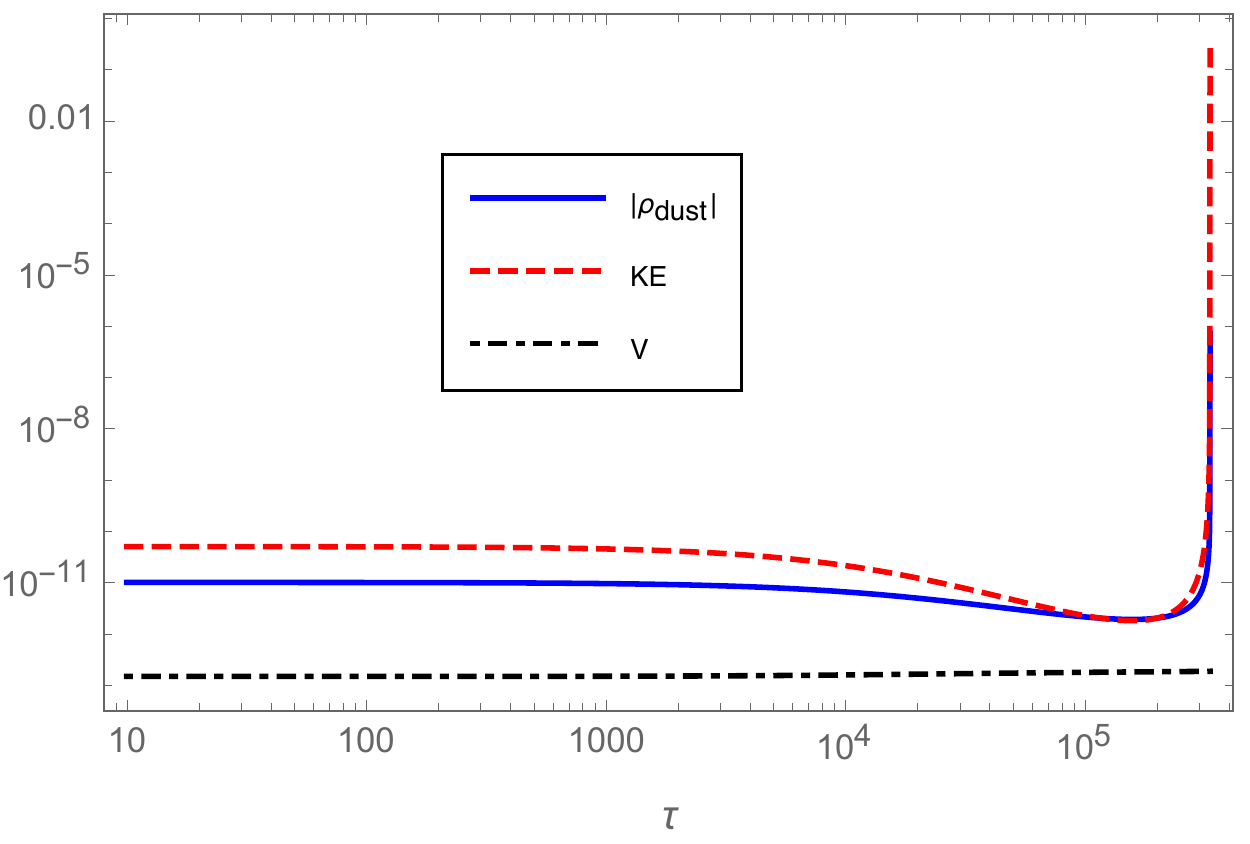}
\includegraphics[width=8cm]{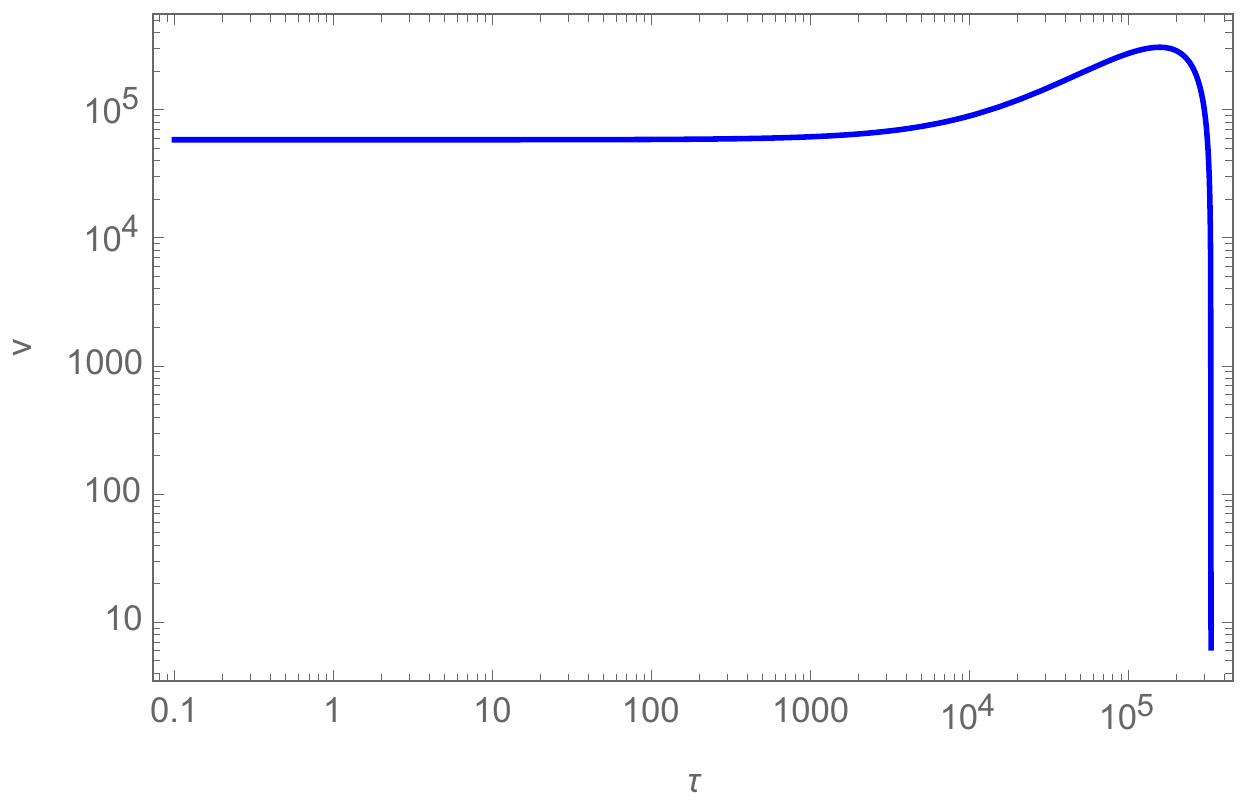}
}
\caption{With the initial conditions given by   (\ref{initial4}),  in the left panel, we show the behavior of the energy density of the dust field and the kinetic/potential energy of the scalar field. The volume of the universe depicted in the right panel indicates that, instead of the slow-roll phase, the recollapse of the universe takes  place  when the magnitude of the negative dust energy density is large enough.}

\label{f4}
\end{figure}

In addition to positive dust energy density,  we also obtained numerical  solutions  with a negative dust energy density for the Brown-Kucha\v{r}  dust. Such negative energy densities though not common in cosmology can occur in situations such as phantom fields \cite{phantom1,phantom2} which have been also considered earlier in the relational formalism \cite{t2006}. It is obvious that  the slow-roll inflation can only take place when the  the magnitude of the negative dust energy density is small enough. If the dust field with negative energy density  plays the dominant role, there will be a recollapse  of the universe leading to a big-crunch singularity.  These results are explicitly shown in Figs. \ref{f3}-\ref{f4}. In Fig. \ref{f3}, the initial conditions are chosen to be 
\bqn
\lb{initial3}
A_0&=&1.50\times10^3, \quad \overline \Phi_0=0.540, \quad  \dot{\overline \Phi}_0=10^{-5}, \nb\\
\rho_{0,\Phi}&=&5.01\times10^{-11}, \quad \rho_{0,\mathrm{dust}}=-10^{-13},\quad  \overline {\mathcal E}^{\mathrm{dust}}=-5.81\times10^{-9}.
\eqn
From Fig. \ref{f3}, we  find the slow-roll inflation can still take place  for  the negative dust energy density as long as the scalar field is the dominant component of the universe at all times. The total number of the inflationary e-foldings turns out to be $68.8$ which is larger than the one from the case of the positive dust energy density. This can be understood from noting that in presence of negative energy density, Hubble rate decreases and hence the Klein-Gordon field experiences less Hubble friction while climbing up to the turnaround point. If the initial conditions are chosen such that the field rolls down the potential, the  dust field with negative energy density yields a lower number of e-foldings than model without dust or with positive dust energy density when starting from the same value of scalar field at the onset of inflation. In such a case, dust with positive energy density increases the number of e-foldings because of an increase in Hubble friction. 

On the other hand, the most dramatic situation occurs in the Brown-Kucha\v{r} model where  the negative energy density causes a recollapse of the universe resulting in a big crunch singularity. 
An example is the following choice of initial conditions:
\bqn
\lb{initial4}
A_0&=&1.50\times10^3, \quad \overline{\Phi}_0=0.540, \quad  \dot{\overline{\Phi}}_0=10^{-5}, \nb\\
\rho_{0,\Phi}&=&5.01\times10^{-11}, \quad \rho_{0,\mathrm{dust}}=-10^{-11},\quad  \overline {\mathcal E}^{\mathrm{dust}}=-5.81\times10^{-7}.
\eqn
The numerical solutions under the above initial conditions are shown in Fig. \ref{f4}.  In this case, the magnitudes of the energy densities of the dust and scalar fields become equal to each other at the  recollapse of the universe. The volume of the universe shrinks rapidly after $\tau=2\times10^5$ and a big crunch singularity is reached with a divergence in the energy densities. Such cases provide a constraint on appropriate values of the initial dust energy density. In order to extract physically  meaningful predictions from the  Brown-Kucha\v r dust model, one should avoid the singular solutions as shown in Fig. \ref{f4} by choosing  the absolute value of the dust energy density comparatively smaller than the scalar field.

In summary, we find that the initial energy density of the dust fields can effectively change  the number of inflationary e-folds. In the case when the inflaton initially climbs up the inflationary potential before slow-rolling down, if dust energy density is positive it has the effect of reducing  inflationary e-foldings. On the contrary,  the negative dust energy density can prolong the inflationary phase if the dust energy density plays a subdominant role all the times. Meanwhile, a too large negative dust energy can lead to the recollapse of the universe resulting in a  big-crunch singularity. Apart from the initial conditions (\ref{initial1}), (\ref{initial3}) and (\ref{initial4}), we also obtained numerical solutions with the initial conditions in which the inflaton is initially rolling down the inflationary potential. In this case, positive dust energy density can lead to a larger period of the inflationary phase while the negative dust energy density has an opposite effect.

\section{Manifestly gauge-invariant perturbations in a spatially-flat FLRW universe with dust reference fields}
\label{sec:Perturbation}
\renewcommand{\theequation}{4.\arabic{equation}}\setcounter{equation}{0}
In this section, we investigate  linear perturbations in the relational formalism with Gaussian and the Brown-Kucha\v{r} dust. In the previous section, we found that the corrections to the background dynamics due to the dust fields can be quantified by formally adding  the dust contributions to the Friedmann and Raychaudhuri equations. Now we would like to understand the impact of the dust fields on the cosmological perturbations via an analysis of their modifications to the Mukhanov-Sasaki equation in above two models. Here, it is important to note that this Mukhanov-Sasaki equation was derived choosing a set of physical variables on the reduced phase space that involve the analogue of the Mukhanov-Sasaki variable in standard cosmological perturbation theory that can be constructed in the relational formalism using the observable map with dust reference fields. In the following, we will focus on scalar perturbations only. First we will use Hamilton's equations from the last section to derive the resulting Mukhanov-Sasaki equations for the Gaussian and Brown-Kucha\v{r} models. Finally,  we numerically understand the way coefficients of various perturbation terms in the modification to the above Mukhanov-Sasaki equation behave during inflation sourced by a Starobinsky potential. We find that these coefficients decay rapidly during inflation.

\subsection{Linear perturbations of the spatially-flat FLRW universe} 

In the relational formalism with Gaussian and Brown-Kucha\v{r} dust fields, there are seven physical degrees of freedom in configuration space. Six of these arise from $Q_{ij}$ and one from the scalar field $\Phi$. As a result, the perturbations of the metric and scalar fields contain altogether three scalar modes, two vector modes and two tensor modes. In the following, we will focus on the scalar modes, especially, the analogue of the comoving curvature perturbations in the standard perturbation theory of cosmology. Our strategy is as follows.  We will first find Hamilton's equations for the perturbations in the reduced phase space and then project these equations onto the scalar sector. After finding the equations of motion for each individual  manifestly gauge-invariant scalar mode, we choose a set of independent variables on the reduced phase space and derive the corresponding Mukhanov-Sasaki equation for the Mukhanov-Sasaki variable defined using (\ref{4b8}). We will present the derivations only for the Brown-Kucha\v{r} model  in detail since the equations of motion for the Gaussian dust model can be understood as a special case of the Brown-Kucha\v{r} model when we specialize lapse and shift to for the reduced phase space to $N=1$ and $N^i=0$.

In the relational formalism with Brown-Kucha\v{r} dust reference fields, the lapse function and the shift vector can be expressed in terms of the canonical  variables on the reduced phase space, therefore, we only needs to consider the physical degrees of freedom, that is the perturbations of the metric and the scalar field around the FLRW background. This means that the perturbations can be parametrized as 
\bqn
\lb{4b1}
Q_{ij}&=&\overline Q_{ij}+\delta Q_{ij},\\
P^{ij}&=&\overline P^{ij}+\delta P^{ij},\\
\Phi&=&\overline \Phi+\delta \Phi,\\
\Pi_\Phi&=&\overline \Pi_\Phi+\delta \Pi_\Phi.
\eqn
From Hamilton's equations of the manifestly gauge-invariant canonical variables in  (\ref{bkeom1}), it is straightforward to find the evolution equations of their perturbations, which turn out to be \cite{ghtw2010I,ghtw2010II}
\bqn
\lb{4b2}
\delta \dot \Phi&=&\frac{\lambda_\varphi\delta \Pi_\Phi}{A^{3/2}}-\frac{\lambda_\varphi\overline \Pi_\Phi}{2A^{5/2}}\delta Q_{ii},\\
\delta \dot \Pi_\Phi&=&\frac{1}{2\lambda_\varphi}\left(-\frac{\sqrt{A}}{2}V_{,\Phi}\delta Q_{ii}-A^{3/2}V_{,\Phi \Phi}\delta \Phi+2\sqrt{A}\Delta \delta \Phi\right)+\overline \Pi_\Phi\partial_i \delta N^i,\\
\delta \dot Q_{ij}&=&\frac{2}{\sqrt A}\overline G_{ijmn}\left(A \delta P^{mn}+\frac{\mathcal P}{2}\delta^{mk}\delta^{nl}\delta Q_{kl}\right)+2 \partial_{(i}\delta N_{j)}, \\
\lb{4b2a}
\delta \dot P^{ij}&=&-\frac{2\mathcal P}{A}\overline G^{ijmn}\partial_{(m} \delta N_{n)}-\frac{\mathcal P}{\sqrt{A}}\overline G_{ijmn} \delta P^{mn}-\frac{1}{\sqrt{A}}{{\overline G}_{ij}}^{~~mn}\delta R_{mn}-\frac{1}{\sqrt{A}}\left(\frac{5\mathcal P^2}{4A}+\frac{\kappa}{2}\overline p_s\right)\delta Q_{ij}\nb\\
&&+\frac{\delta_{ij}}{\sqrt{A}}\left(\frac{3\mathcal P^2}{8A}-\frac{\kappa \overline \rho_s}{4}\right)\delta Q_{kk}+\delta_{ij}\frac{\kappa}{4\lambda_\varphi}\left(\frac{2\lambda^2_\varphi\overline \Pi_\Phi}{A^{5/2}}\delta \Pi_\Phi-\sqrt{A}V_{,\Phi}\delta \Phi\right) ~.
\eqn
Here  we have used the fact that $\overline Q_{ij}=A\delta_{ij}$, $\overline{P}^{ij}=\mathcal P\delta^{ij}$, $\overline N=1$, $\overline N^i=0$ and $\delta N=0$, the latter comes from the fact that in the Brown-Kucha\v{r} model the perturbations of $N$ have non-trivial contribution only from second order on. 
Besides, in the above formulae, $\delta Q_{ii}=\delta^{ij}\delta Q_{ij}$ and  
\bqn
\lb{4b3}
\overline G_{ijmn}&=&\frac{1}{2}\left(\delta_{im}\delta_{jn}+\delta_{in}\delta_{jm}-\delta_{ij}\delta_{mn}\right), \nb \\
\overline G^{-1}_{ijmn}&=&\frac{1}{2}\left(\delta_{im}\delta_{jn}+\delta_{in}\delta_{jm}\right)-\delta_{ij}\delta_{mn},\nb\\
\overline G^{ijmn}&:=&\overline G_{ijmn}.
\eqn
Now as usually done in cosmological perturbation theory we can decompose the metric perturbations as well as their momenta into scalar, vector and tensor (SVT) modes as  
\bqn
\lb{4b4}
\delta Q_{ij}&=&2A\left(\psi \delta_{ij}+E,_{<ij>}+F_{(i,j)}+\frac{1}{2}h^{TT}_{ij}\right),\\
\delta P^{ij}&=&2\mathcal P\left(p_\psi \delta^{ij}+p_E^{,<ij>}+p_F^{(i,j)}+\frac{1}{2}p^{ij}_h\right),
\eqn
where we used that any symmetric rank 2 tensor $T_{ij}$ that describes perturbations around a flat FLRW metric  can be decomposed into two scalar components, two transversal vector components and two transversal traceless tensor components as
\bq
\lb{4a3}
T_{ij}=\frac{1}{3}\delta_{ij}T+2\partial_{<i}\partial_{j>}T_S+2\partial_{(i}T_{j)}+T^{TT}_{ij},
\eq
where $\partial_{<i}\partial_{j>}:=\partial_{(i}\partial_{j)}-\frac{1}{3}\delta_{ij}A\Delta$ with $\Delta:=\delta^{jk}\partial_j\partial_k$. Furthermore, we introduced the trace $T:=Q^{jk}T_{jk}$, $T_S$ is the longitudinal scalar component, $T_i$ denotes the  longitudinal transversal components, and $T^{TT}_{ij}$ are the transversal traceless components whose explicit forms can for instance be found in \cite{Baumann,gh2018}. The individual manifestly gauge-invariant scalar perturbations relevant for us  are explicitly given by
\bqn
\lb{4b5}
\psi&=&\frac{\delta^{ij}\delta Q_{ij}}{6A}, \quad \quad p_\psi=\frac{\delta_{ij} \delta P^{ij}}{6\mathcal P},\\
E&=&\frac{3}{4A}\Delta^{-2}\partial^{<i}\partial^{j>}\delta Q_{ij},\\
p_E&=&\frac{3}{4\mathcal P}\Delta^{-2}\partial_{<i}\partial_{j>}\delta P^{ij}.
\eqn
The reason why such a SVT decomposition is of advantage is because in linear order the equations of the  scalar, vector and tensor sector decouple and can thus be solved independently. 

Using  (\ref{4b2})-(\ref{4b2a}), we  can derive the equations of the motion of the manifestly gauge-invariant individual scalar perturbations, which turn out to be
\bqn
\lb{scalarmode}
\delta \dot \Phi&=&\frac{\lambda_\varphi}{A^{3/2}}\left(\delta \Pi_\Phi -3\overline \Pi_\Phi \psi \right),\\
\delta \dot \Pi_\Phi&=&\frac{1}{2\lambda_\varphi}\left(-3 A^{3/2}V_{,\Phi}\psi-A^{3/2}V_{,\Phi \Phi}\delta \Phi+2\sqrt{A}\Delta \delta \Phi\right)+\frac{\overline \Pi_\Phi}{A \overline C}\Delta \delta \mathcal E^{\mathrm{dust}}_{\parallelsum},\\
\dot \psi&=&-\frac{\mathcal P}{\sqrt A}p_\psi+\frac{\mathcal P}{2\sqrt A}\psi+\frac{\Delta \delta \mathcal E^{\mathrm{dust}}_{\parallelsum}}{3A\overline C},\\
\lb{scalare}
\dot E&=&\frac{2\mathcal P}{\sqrt A} \left(E+p_E \right)+\frac{ \delta \mathcal E^{\mathrm{dust}}_{\parallelsum}}{A\overline C},\\
\dot p_\psi&=&\frac{\mathcal P}{4\sqrt A}p_\psi-\frac{\mathcal P\psi}{8\sqrt A}+\frac{\kappa \lambda_\varphi }{4\mathcal PA^{5/2}}\left(\overline \Pi_\Phi \delta \Pi_\Phi-\overline \Pi^2_\Phi p_\psi\right)-\frac{1}{3\mathcal P\sqrt A}\left(\Delta \psi-\frac{1}{3}\Delta^2 E\right)\nb\\
&&+\frac{\kappa \sqrt A}{8 \mathcal P\lambda_\varphi}\left(2p_\psi V-V_{,\Phi}\delta \Phi \right)-\frac{\kappa \sqrt A \psi}{4 \mathcal P}\left(3 \rho_\Phi+2 p_\Phi\right)+\frac{1}{6A\overline C}\Delta \delta \mathcal E^{\mathrm{dust}}_{\parallelsum},\\
\lb{scalarmode1}
\dot p_E&=&\frac{1}{2\mathcal P\sqrt A}\left(\psi-\frac{1}{3}\Delta E\right)-\left(\frac{\mathcal P}{2\sqrt A}+\frac{\kappa\lambda_\varphi \overline{\Pi}^2_\Phi}{2\mathcal PA^{5/2}}+\frac{\kappa}{2\mathcal PA}\overline{\mathcal E}^\mathrm{dust}\right)\left(E+p_E\right)-\frac{ \delta \mathcal E^{\mathrm{dust}}_{\parallelsum}}{A \overline C} .
\eqn
Here $\rho_\Phi$ and $p_\Phi$ are the background energy density and pressure given by (\ref{3a12}), and $\delta \mathcal E^{\mathrm{dust}}_{\parallelsum}$ is  the longitudinal scalar projection of the dust contributions to the spatial diffeomorphism constraint  and $\delta \mathcal E^{\mathrm{dust}}$  the perturbation of the dust contribution to the Hamiltonian constraint  defined as
\bqn
\lb{4b6}
&&\delta \mathcal E^{\mathrm{dust}}_{\parallelsum}\equiv-\Delta^{-1}\partial_i \delta C_i/\kappa \equiv \overline C \Delta^{-1}\partial_i\delta N_i/\kappa=-\overline \Pi_\Phi\delta \Phi+\frac{4A\mathcal P}{\kappa }\Bigg\{\frac{2}{3}\Delta \left(E+p_E\right)+p_\psi-\frac{\psi}{2}\Bigg\}, \\
&&\delta \mathcal E^{\mathrm{dust}}\equiv -\delta C/\kappa=-\frac{4\sqrt A}{\kappa}\Delta\left(\psi-\frac{1}{3}\Delta E\right)-\frac{A^{3/2}}{2\lambda_\varphi}\left(V_{,\Phi}\delta \Phi +3\psi V\right)\nb\\
&&~~~~~~~~~~~~~~~~~~~~~~-\frac{\lambda_\varphi\overline\Pi_\Phi}{A^{3/2}}\left(\delta \Pi_\Phi-\frac{3}{2}\overline\Pi_\Phi \psi\right)+\frac{3\sqrt A \mathcal P^2}{2\kappa}\left(\psi+4p_\psi\right) ~.
\label{4b6b}
\eqn
Comparing the corresponding equations of motion of the perturbations in the Gaussian dust model \cite{laura} one realizes that these can then be obtained from above equations  by simply removing the terms proportional to $\delta \mathcal E^{\mathrm{dust}}_{\parallelsum}$ in  (\ref{scalarmode})-(\ref{scalarmode1}).

Now we can  proceed to find the Mukhanov-Sasaki equation. The strategy we follow here is as follows: In the conventional approach without the dust reference fields the perturbation of the scalar field $\delta\varphi$ is not gauge-invariant as this is also true for $\psi,E$ in contrast to our $\delta\Phi,\psi,E$ here. Therefore, in the conventional case one chooses a specific combination of $\delta\varphi,\psi,E$ and background quantities in order to obtain the conventional Mukhanov-Sasaki variable that is invariant under linearized diffeomorphisms. In order to compare our approach to the conventional one and thus be able to analyze the corrections of the reference fields, we define a  Mukhanov-Sasaki variable in terms of our already gauge-invariant perturbations in the following way  \cite{Langlois_1994}
\bq
\lb{4b8}
Q=\delta \Phi+  Z \left(\psi-\frac{\Delta}{3}E\right),
\eq
where $Z=2\lambda_\varphi \frac{\overline\Pi_\Phi}{A\mathcal P}$. In contrast to the conventional treatment note that in our case each individual term in (\ref{4b8}) is already manifestly gauge-invariant. 
For both the Brown-Kucha\v{r} and Gaussian dust model, a straightforward calculations using the corresponding Hamilton's equations results in the following equation of motion for $Q$:
\bq
\lb{4b9}
\ddot Q+\frac{3}{2}\frac{\dot A}{A}\dot Q-\left(\frac{\Delta}{A}+\frac{3}{2}\frac{\dot A}{A}\frac{\dot Z}{Z}+\frac{\ddot Z}{Z}\right)Q=F^{\rm BK/G}_\mathrm{dust}.
\eq
where $F^{\rm BK/G}_\mathrm{dust}$ is the additional term accounting for the contributions from the dust reference fields either in the Brown-Kucha\v{r} or Gaussian dust model.  In order to find the explicit form of $F^{\rm BK/G}_\mathrm{dust}$, one can start from the equations of motion of the scalar perturbations given in (\ref{scalarmode})-(\ref{scalarmode1}) which are valid for the Brown-Kucha\v r dust fields. When dropping the terms proportional to  $\delta \mathcal E^{\mathrm{dust}}_{\parallelsum}$ in (\ref{scalarmode})-(\ref{scalarmode1}), one can recover the equations of motion of the scalar modes in the Gaussian dust model. Based on the definition of the Mukhanov-Sasaki variable in (\ref{4b8}), it is straightforward to compute the left-hand side of the equation (\ref{4b9}), the result of which gives the correction term which explicitly takes the form  \cite{laura}
\bqn
\lb{4b11}
&&F^{\rm BK/G}_\mathrm{dust}=\left(-\frac{3\kappa\lambda_\varphi \overline{\Pi}_\Phi}{2 A^3}+\frac{\kappa^2\lambda^2_\varphi\overline{\Pi}^3_\Phi}{2A^5\mathcal P^2}+\frac{\kappa^2\lambda_\varphi\overline{\Pi}_\Phi \overline{\mathcal E}^\mathrm{dust}}{2A^{7/2}\mathcal P^2}+\frac{\kappa V_{,\Phi}}{2A\mathcal P}\right)\delta \mathcal E^{\mathrm{dust}}_{\parallelsum}+\frac{\kappa \lambda_\varphi\overline{\Pi}_\Phi}{2A^{5/2}\mathcal P}\delta \mathcal E^{\mathrm{dust}}\nb\\
&&+\overline{\mathcal E}^{\mathrm{dust}}\Bigg[-\frac{\kappa\lambda_\varphi\overline{\Pi}_\Phi}{2A^{5/2}\mathcal P}\Delta E-\frac{3\kappa Q}{4A^{3/2}}+ \delta \Phi\left(\frac{3\kappa}{2A^{3/2}}-\frac{\kappa^2\lambda_\varphi \overline{\Pi}^2_\Phi}{2A^{7/2}\mathcal P^2}-\frac{\kappa\sqrt A V_{,\Phi}}{2\lambda_\varphi \mathcal P\overline{\Pi}_\Phi}-\frac{\kappa^2\overline{\mathcal E}^{\mathrm{dust}}}{2A^2\mathcal P^2}\right)\Bigg]~.
\eqn
Note that  the dust correction takes the same form in the Brown-Kucha\v r  and the Gaussian dust models. The extra terms proportional to  $\delta \mathcal E^{\mathrm{dust}}_{\parallelsum}$ in (\ref{scalarmode})-(\ref{scalarmode1}) in the Brown-Kucha\v r dust model do not give rise to new correction terms in  $F^{\rm BK}_\mathrm{dust}$ because they cancel each other. 
However, even though the form of the  Mukhanov-Sasaki  equation is identical in the  Brown-Kucha\v r  and Gaussian dust models,  the evolution of the  Mukhanov-Sasaki  variables will be different. The reason is because $\delta {\cal E}^{\mathrm{dust}}$ is {\it{not}} a constant of motion in the Gaussian dust model unless $\delta {\cal E}^{\mathrm{dust}}_{\parallelsum}$ vanishes. That this behavior is completely consistent with the perturbed constraint algebra will be discussed at the end of this section.
Further,  the Mukhanov-Sasaki  equation (\ref{4b9}) is not in a closed form as $F^{\rm BK/G}_\mathrm{dust}$ also depends on $E$ and $\delta \Phi$. The latter two scalar modes are explicitly dependent on $\delta \mathcal E^{\mathrm{dust}}_{\parallelsum}$ in the Brown-Kucha\v r dust model as shown in (\ref{scalarmode}) and (\ref{scalare}). Thus, $Q$ behaves differently in the Brown-Kucha\v r  and the Gaussian dust models even when one chooses same initial conditions for the perturbations. Our following discussion of numerical results confirms that the Mukhanov-Sasaki variable has a different evolution for the two dust models. Moreover, it is also worthwhile to point out that the explicit expression of these correction terms are tied to the chosen set of independent physical variables on the reduced phase space that involve the form of the Mukhanov-Sasaki variable defined using (\ref{4b8}).  Besides, the third term in  (\ref{ham}) does not contribute to the dynamics of the linear perturbations. Instead, it  only affects the higher order perturbations.

\begin{figure}[H]
{
\includegraphics[width=8cm]{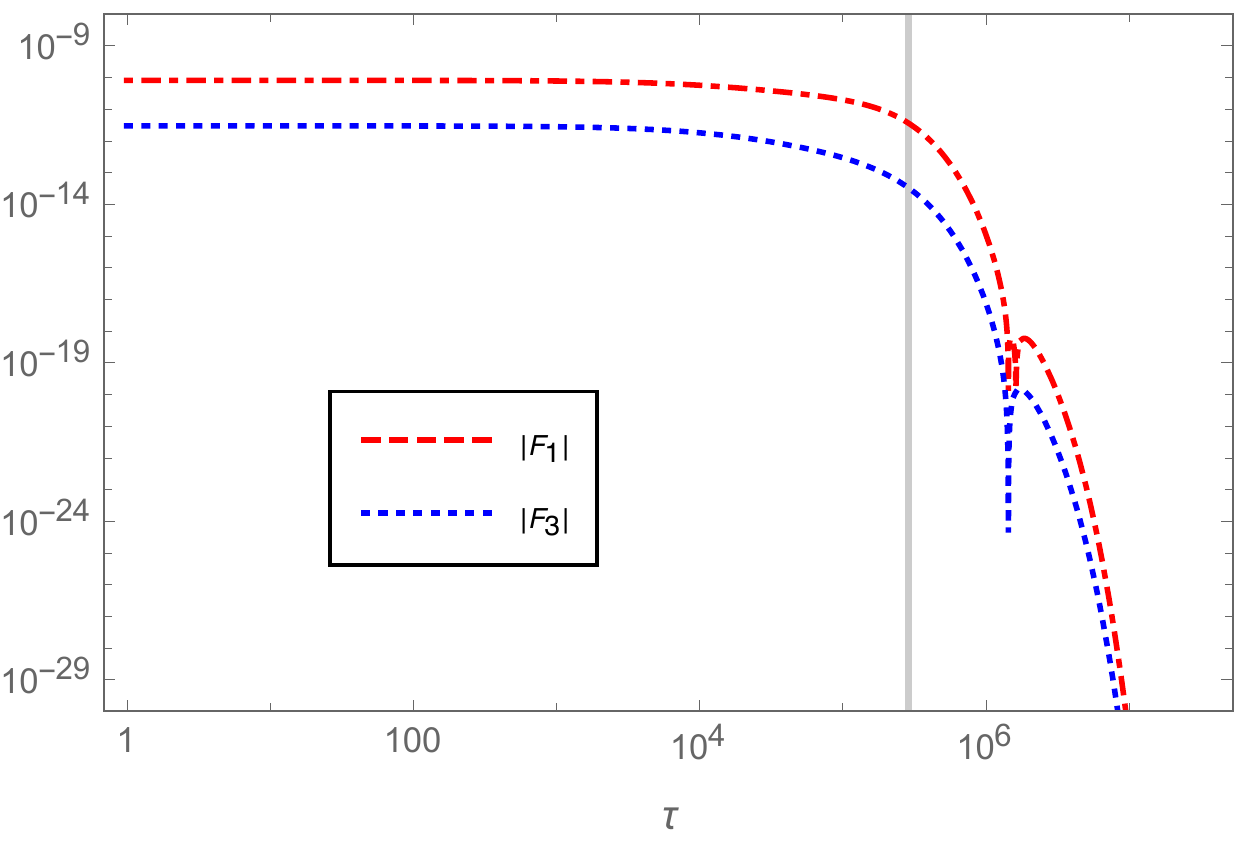}
\includegraphics[width=8cm]{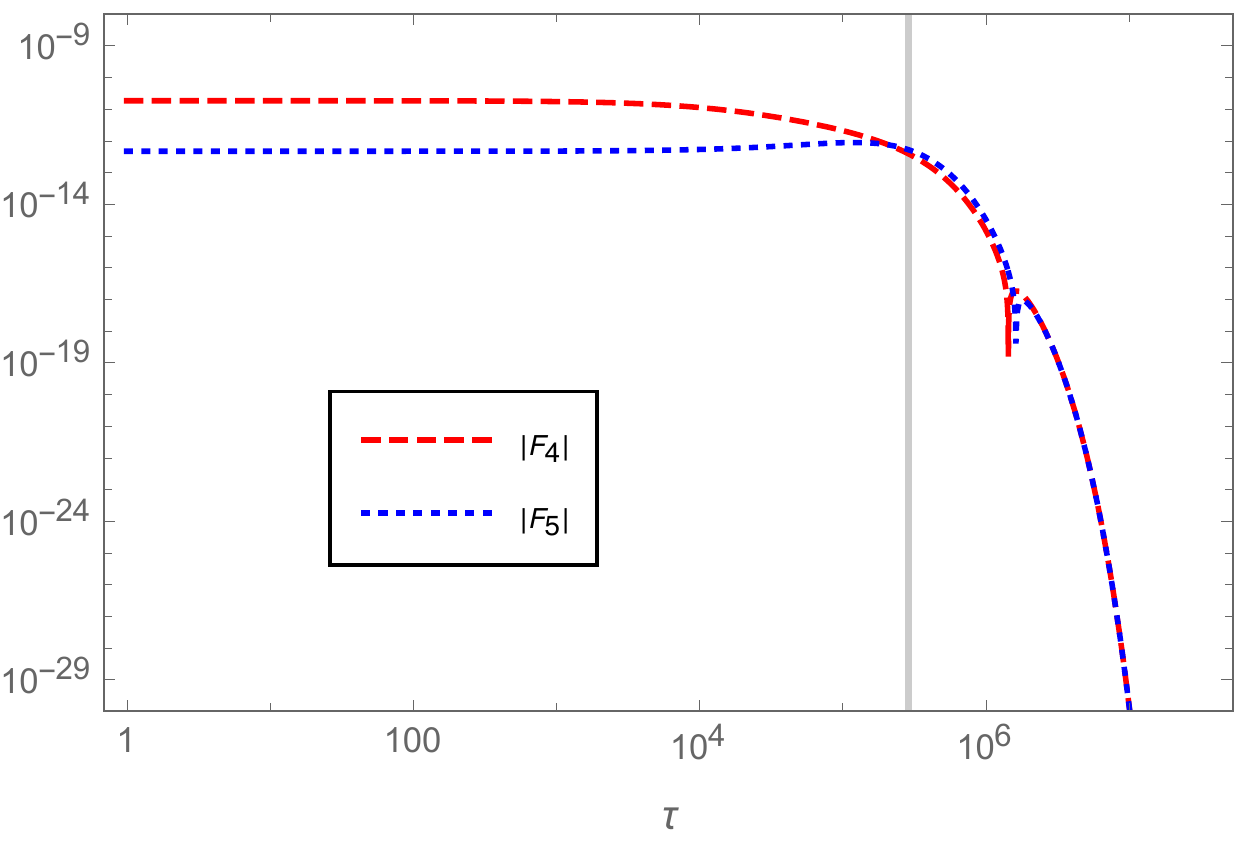}
\includegraphics[width=8cm]{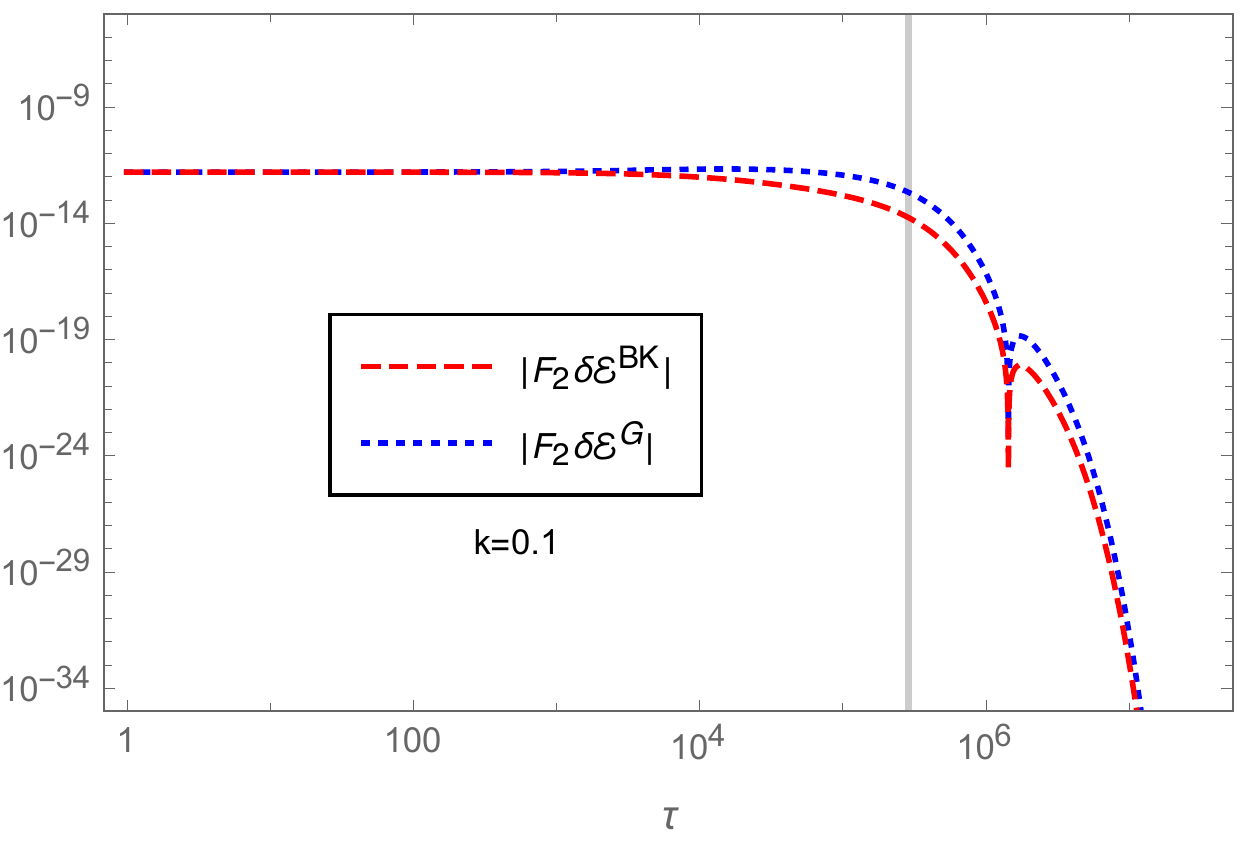}
\includegraphics[width=8cm]{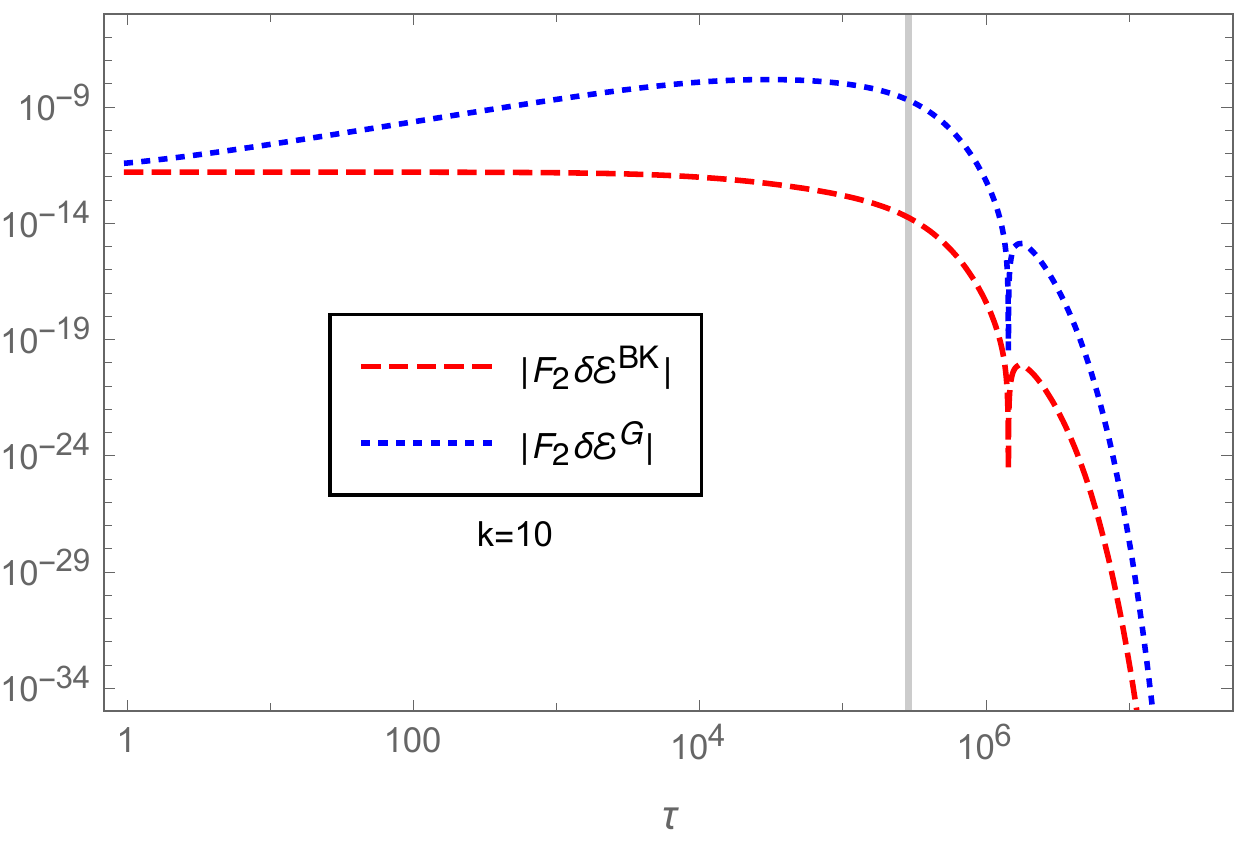}
}
\caption{With the same initial conditions of the background given in (\ref{initial1}), the evolution of the background-dependent coefficients in (\ref{4b13}) are shown in  the preinflationary and the slow-roll phases. The vertical lines in the subfigures mark the onset of inflation. All the coefficients decay exponentially in the slow-roll phase. In particular, since $ \delta \mathcal E^{\mathrm{dust}}$ is in general not a constant of motion in the Gaussian dust model, we show the behavior of $F_2 \delta \mathcal E^{\mathrm{dust}}$ in two dust models for the particular wavenumber $k=0.1$ and $k=10$. In the figure, $\delta \cal E^\mathrm{G}$ denotes $\delta \cal E^\mathrm{dust}$ in the Gaussian dust model and  $\delta \cal E^\mathrm{BK}$ in the Brown-Kucha\v r dust model. In the figure, the $k$ dependence of $\delta \mathcal E^{\mathrm{dust}}$ is omitted to keep our notation compact.  }
\label{f6}
\end{figure}

\begin{figure}[H]
{
\includegraphics[width=8cm]{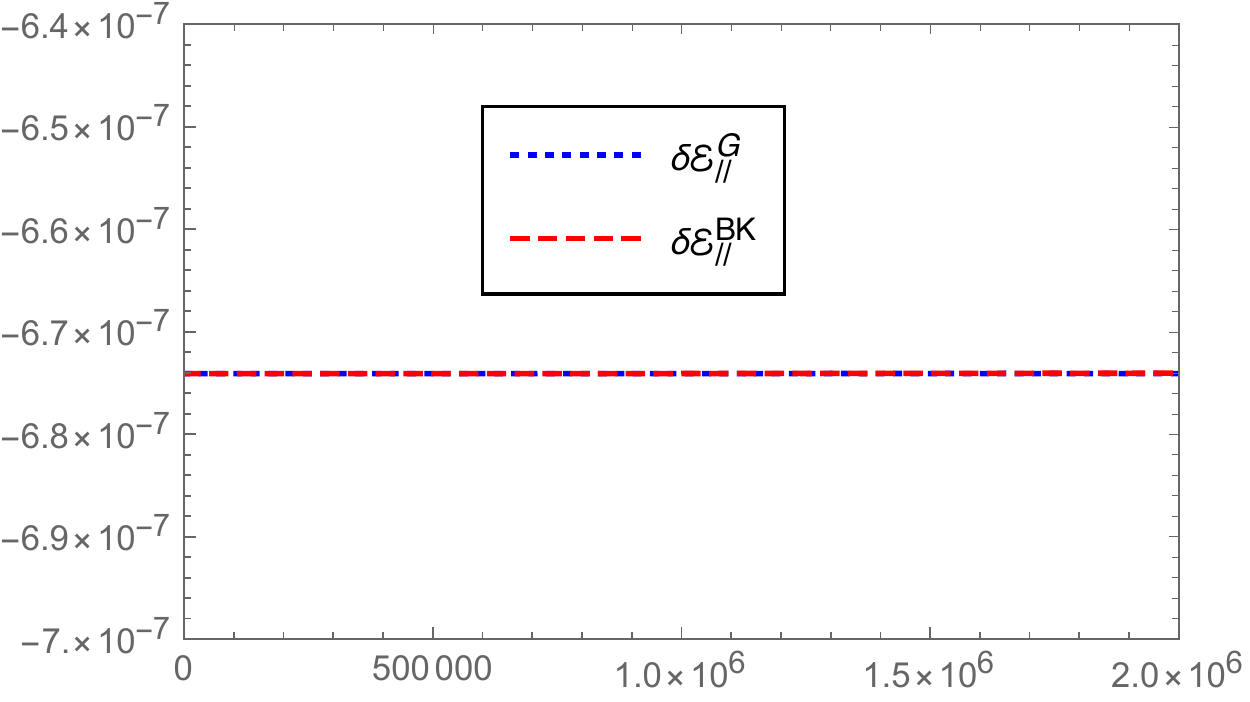}
\includegraphics[width=8cm]{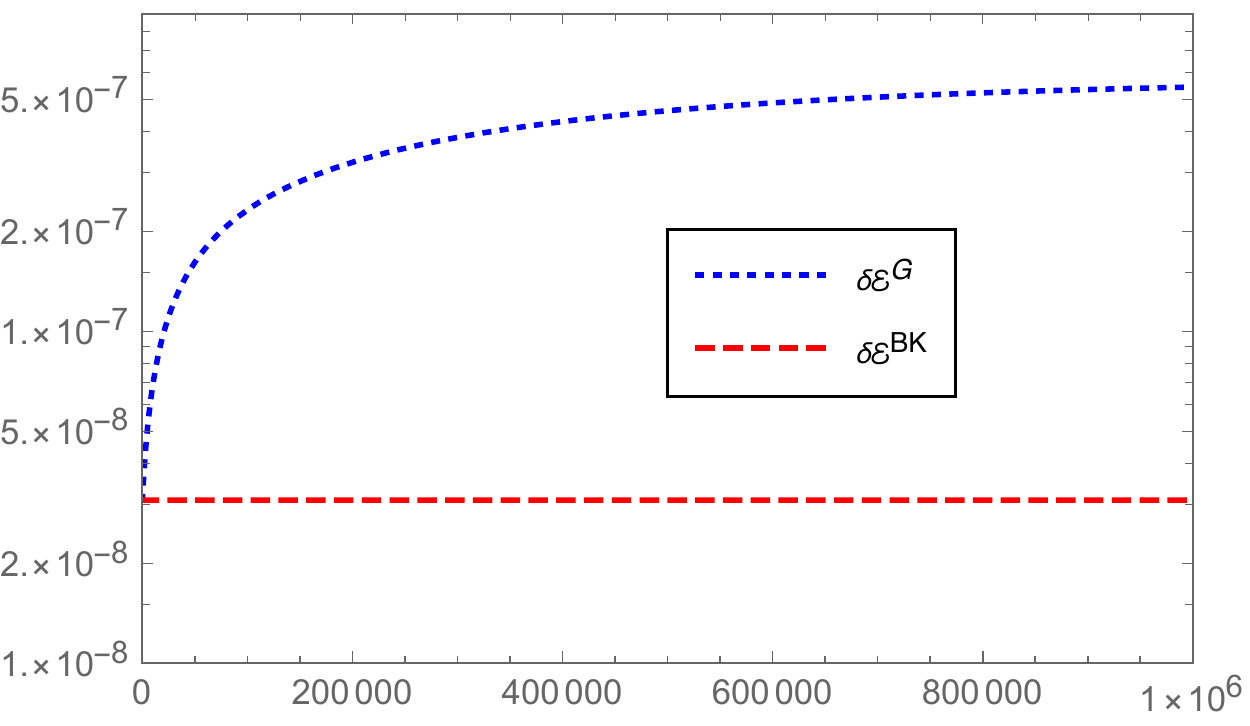}
\includegraphics[width=8cm]{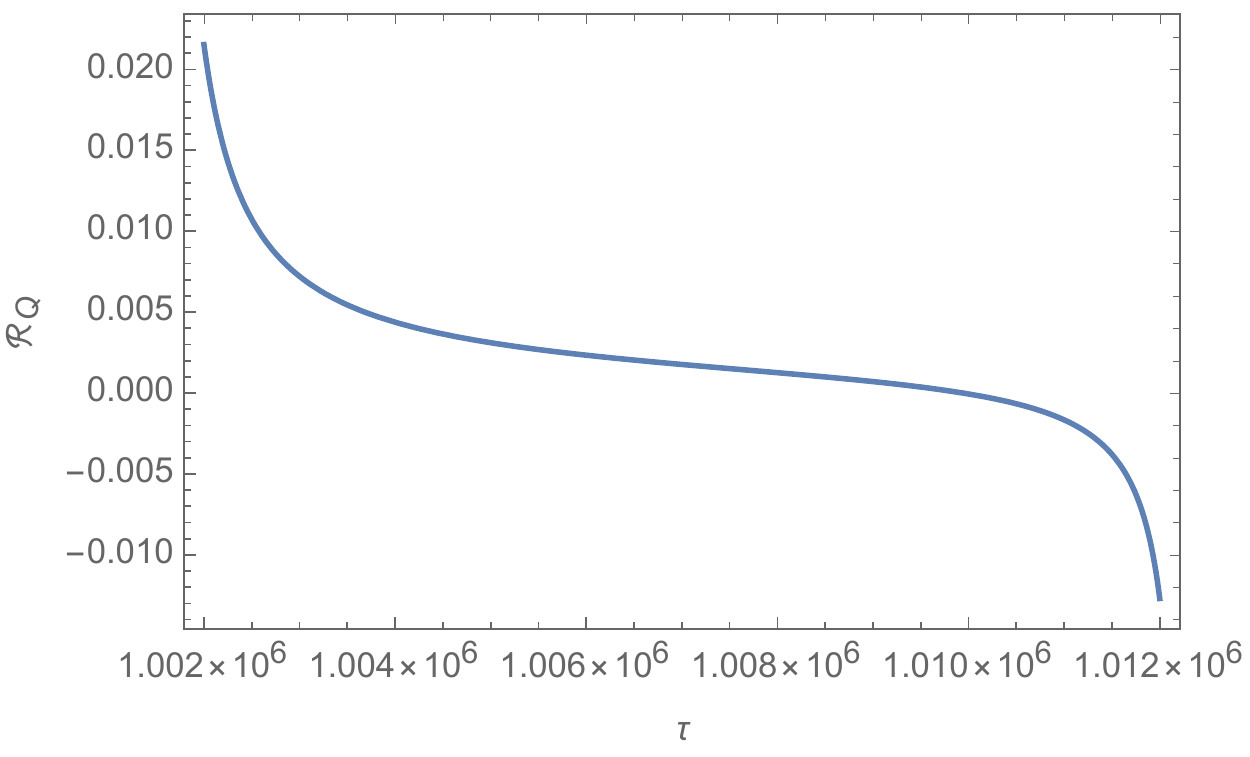}
\includegraphics[width=8cm]{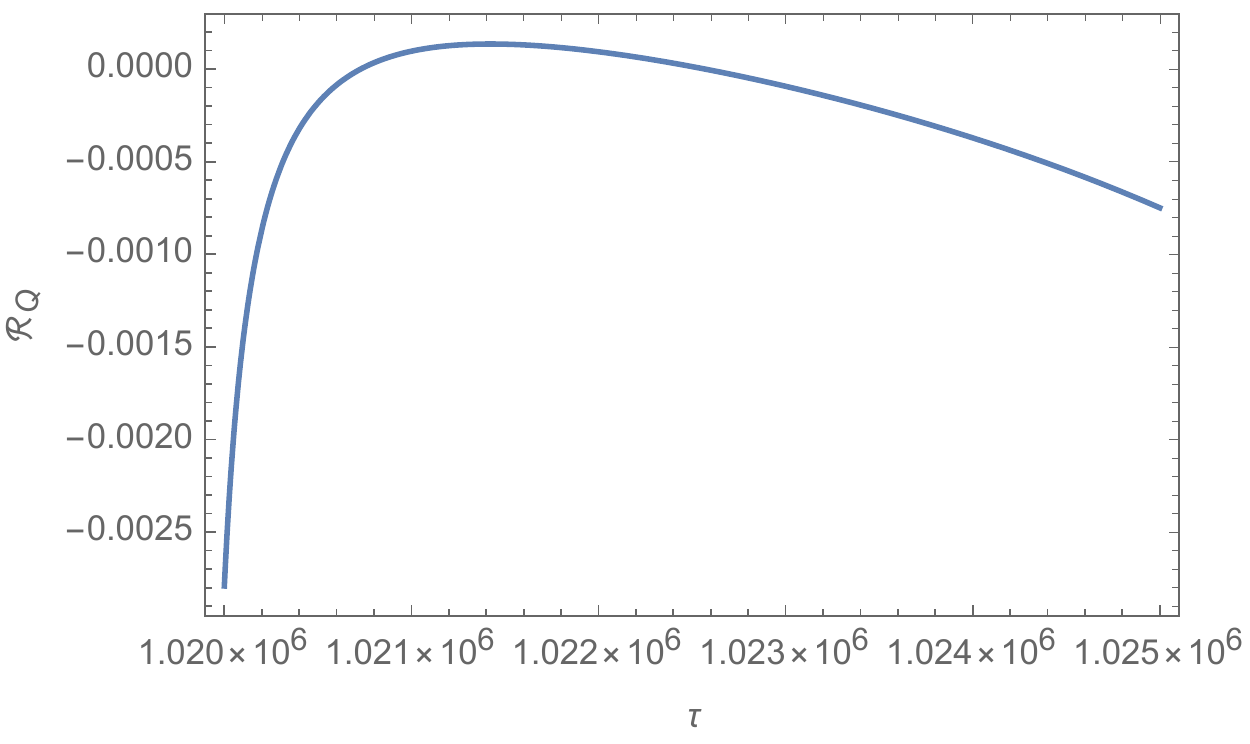}
}
\caption{With the initial conditions for the background given by (\ref{initial1}), the coupled equations of motion in (\ref{scalarmode})-(\ref{scalarmode1}) are evolved from $t=0$ for the particular comoving wavenumber $k=0.1$ with all the initial magnitudes of the scalar modes set to $10^{-6}$. In the top panels, we compare the behavior of  $\delta \mathcal E^{\mathrm{dust}}_{\parallelsum}$ and $ \delta \mathcal E^{\mathrm{dust}}$ in the Gaussian and Brown-Kucha\v r dust models. In the bottom panels, the relative difference of the Mukhanov-Sasaki variable between two dust models are depicted near $t=10^6$. In the figure, the $k$ dependence of $\mathcal R_Q$ is omitted for a simpler notation.}
\label{f5}
\end{figure}

An important question is about the impact of dust correction terms $F^\mathrm{BK/G}_\mathrm{dust}$. To answer this question completely we need to compute the power spectrum of the perturbations by solving the coupled system of equations of motion for all gauge-invariant perturbations with suitable initial conditions. In our future work, we will undertake this task by a detailed phenomenological investigation on implications from differences in two models. However, a glimpse on the behavior of these terms can be obtained by understanding the way the coefficients of different perturbation terms in $F^\mathrm{BK}_\mathrm{dust}$ and $F^\mathrm{G}_\mathrm{dust}$ behave. For this, let us introduce the background quantities $F_i$ with $i=1,..,5$ to denote the coefficients of the perturbations that only depend on the background variables, where
\bqn
{F}_1 &=&Z\left( \nonumber -\frac{3\kappa\lambda_\varphi \overline{\Pi}_\Phi}{2 A^3}+\frac{\kappa^2\lambda^2_\varphi\overline{\Pi}^3_\Phi}{2A^5\mathcal P^2}+\frac{\kappa^2\lambda_\varphi\overline{\Pi}_\Phi \overline{\mathcal E}^\mathrm{dust}}{2A^{7/2}\mathcal P^2}+\frac{\kappa V_{,\Phi}}{2A\mathcal P}\right), 
{F}_2= \nonumber \frac{\kappa \lambda_\varphi\overline{\Pi}_\Phi Z}{2A^{5/2}\mathcal P}, {F}_3= \nonumber - \frac{\kappa\lambda_\varphi\overline{\Pi}_\Phi Z}{2A^{5/2}\mathcal P} \overline{\mathcal E}^{\mathrm{dust}}, \\ ~~~ {F}_4& =& - \frac{3\kappa Z}{4A^{3/2}}  \overline{\mathcal E}^{\mathrm{dust}}, ~~~ {F}_5= \left(\nonumber     \frac{3\kappa}{2A^{3/2}}-\frac{\kappa^2\lambda_\varphi \overline{\Pi}^2_\Phi}{2A^{7/2}\mathcal P^2}-\frac{\kappa\sqrt A V_{,\Phi}}{2\lambda_\varphi \mathcal P\overline{\Pi}_\Phi}-\frac{\kappa^2\overline{\mathcal E}^{\mathrm{dust}}}{2A^2\mathcal P^2}\right)  Z\overline{\mathcal E}^{\mathrm{dust}} ~.
\eqn 
Using these we find
\bqn
\lb{4b13}
Z F^\mathrm{BK/G}_\mathrm{dust}={F}_1\delta \mathcal E^{\mathrm{dust}}_{\parallelsum}+{F}_2\delta \mathcal E^{\mathrm{dust}}+{F}_3 \Delta E+{F}_4 Q+F_5 \delta \Phi.
\eqn
In the above formula, an additional overall factor of $Z$ is multiplied to remove the singularity at the turnaround point $\dot \Phi=0$ where both $\overline \Pi_\Phi$ and $Z$ vanish. The behavior of the ${F}_i$ coefficients for representative initial conditions leading to inflation is shown in Fig. \ref{f6}. More specifically, choosing the same initial conditions as in Fig. \ref{f5}, we find that all these coefficients quickly decay at the onset of inflation. In particular, as $\delta \mathcal E^{\mathrm{dust}}$ will in general vary over time in the Gaussian dust model, we compare $F_2 \delta \mathcal E^{\mathrm{dust}}$ in the two dust models for wavenumbers $k=0.1$ and $k=10$ (see Fig. \ref{f6}). A comparison of these terms with other background coefficient terms on the left hand side of the Mukhanov-Sasaki equation (\ref{4b9}) shows that during inflation contributions coming from $F^\mathrm{BK/G}_\mathrm{dust}$  quickly become far less significant. If the amplitude of perturbations remains constant or decays during inflation then above results suggest that the role of dust clocks in the Mukhanov-Sasaki equation for the chosen set of independent variables on the reduced phase space involving the the Mukhanov-Sasaki variable defined in (\ref{4b8}) becomes negligible during the inflationary phase.

Although the Mukhanov-Sasaki equation takes the same form in two dust models, since $F^{\rm BK/G}_\mathrm{dust}$ involves other independent scalar modes,  such as $E$ and $\delta \Phi$, one has to solve the equations of motion given in (\ref{scalarmode})-(\ref{scalarmode1}) before we can compare the evolution of Mukhanov-Sasaki variables for the two considered dust clocks. To understand the qualitative behavior of some interesting quantities, we numerically solved the equations of motion  (\ref{scalarmode})-(\ref{scalarmode1}) for one representative wavenumber $k=0.1$  with the initial magnitudes of all the scalar modes set equal to $10^{-6}$ at an initial time $t=0$ and evolved till $t=2.0\times 10^6$. Results for the
perturbations of the dust energy density $\delta \mathcal E^{\mathrm{dust}}$  and dust momentum energy density $\delta \mathcal E^{\mathrm{dust}}_{\parallelsum}$ are presented in Fig. \ref{f5}. From the top left panel, we see that the perturbation of the dust momentum density is a constant of motion in both Gaussian and Brown-Kucha\v r dust models. However, the perturbation of the dust energy density  is  not a constant of motion in the Gaussian dust model unless for a vanishing  $\delta \mathcal E^{\mathrm{dust}}_{\parallelsum}$
.   A straightforward calculation shows that in the Gaussian dust model
\bq
\label{eq:DotE}
\frac{d}{d\tau} {\delta\mathcal E}^{\mathrm G}=\frac{\Delta }{A}\delta \mathcal E^{\mathrm{dust}}_{\parallelsum},
\eq
while in the Brown-Kucha\v r dust model, using the equations of motion (\ref{scalarmode})-(\ref{scalarmode1}), one can find $\frac{d}{d\tau}\delta \mathcal E^{\mathrm{BK}}=0$ which is consistent with the behavior of $\delta \mathcal E^\mathrm{BK}$ in the above figure. 

It is to be noted that these results are consistent with the constraint algebra. Considering full GR the Poisson bracket of two Hamiltonian constraints yields a result that contains the spatial diffeomorphism constraint again. Since in the Gaussian dust model the physical Hamiltonian ${\bf {\rm H}}_{\rm phys}$ is the integral over the dust manifold of the Hamiltonian density $C$, where $C$ is the contribution of all but the dust fields to the Hamiltonian constraint. Given this from the constraint algebra we know that the evolution of the dust energy density $\epsilon=-C/\kappa$ governed by ${\bf {\rm H}}_{\rm phys}$ needs to be proportional to $C_j$, that is the contribution to the spatial diffeomorphism constraint of all but the dust fields. At the level of linearized perturbation theory this carries over to the fact that $\frac{d}{d\tau}\delta {\cal E}^{\mathrm{dust}}$ has to involve $\delta {\cal E}^{\mathrm{dust}}_{\parallelsum}$ where the additional scale factors and partial derivatives are included in (\ref{eq:DotE}) because of the projections we work with as well as due to the inverse metric and partial derivative that are also involved in the corresponding term in the constraint algebra. On the other hand in the Brown-Kucha\v{r} dust model the physical Hamiltonian density is a Kucha\v{r} density and therefore commutes with itself and thus $\frac{d}{d\tau}{\cal E}^{\mathrm{BK}}$ vanishes.

In the bottom panels of the Fig. \ref{f5}, we plot the relative difference in $Q$ from the two dust models for the comoving wavenumber $k=0.1$. The relative difference in the figure is defined as  $\mathcal R_{Q}=\left(Q^{\mathrm{BK}}-Q^{\mathrm G}\right)/Q^{\mathrm G}$, where the $k$ dependence of $\mathcal R_{Q}$ is omitted in our notation to keep it more simple and compact. Since the behavior of $Q$ is oscillatory in nature, the relative difference is found to be oscillatory with a non-vanishing magnitude. In Fig. \ref{f5}, we have shown this behavior for two different time intervals which shows that even for a very small value of $\delta \mathcal E^{\mathrm{dust}}_{\parallelsum}$, the relative difference in $Q$ is significant for the region near $t=10^6$, the magnitude of  $\mathcal R_Q$ can be as high as $2\%$, and this magnitude increases at larger times.

\section{Conclusions and outlook}
\label{sec:Conclusion}
\renewcommand{\theequation}{5.\arabic{equation}}\setcounter{equation}{0}
The aim of this paper is to consider in the relational formalism two sets of reference fields, namely the Brown-Kucha\v{r} dust and the Gaussian dust models and compare their characteristic properties in linear cosmological perturbation theory in the context of the Mukhanov-Sasaki equation resulting from both of the models. For this purpose, we briefly reviewed the framework of the relational formalism in the extended ADM phase space, which in our work consists of gravity, one massive minimally coupled Klein-Gordon scalar field and the dust reference fields. By means of an observable map that can be constructed in the relational formalism once reference fields are chosen for both dust models, the reduced phase space that encodes the physical degrees of freedom in terms of manifestly gauge-invariant Dirac observables was constructed building on earlier works in \cite{ghtw2010I,ghtw2010II,gt2015}. 
An advantage of these reference fields is that  they allow to rewrite the Hamiltonian constraint in a deparametrized form. As a result, the associated physical Hamiltonian that generates the equations of motion of these observables is a constant of motion and can be understood as the energy of the system on the reduced phase space. Hence, at the level of the reduced phase space, GR can be mapped to a conventional Hamiltonian system with a generator of the dynamics that does not vanish on the constraint surface and where all constraints of the system have already been reduced. In the Brown-Kucha\v{r} model lapse $N$ and shift vector $N^i$ are non-constant functions unlike in the case for Gaussian dust where $N=1$ and $N^i = 0$. A comparison between Hamilton's equations in the Brown-Kucha\v{r} and the Gaussian dust models shows that the latter can be obtained from the former by setting the lapse $N$ to unity and the shift vector $N^i$ to vanish on the reduced phase space.

To explore the consequences of dust reference fields on cosmological perturbations,  we first analyzed the reduced phase space of a spatially flat FLRW universe and presented the gauge-invariant Friedmann (\ref{Friedmann}) and Raychaudhuri (\ref{3a9}) equations in the relational formalism. In this symmetry reduced case, we have only one temporal reference dust field because the spatial diffeomorphism constraints are trivially satisfied. As a result the physical Hamiltonians in the Brown-Kucha\v{r} and the Gaussian dust models agree, yielding the same background dynamics. Depending on the classical solutions of GR that one wants to study in the reduced phase space within the Brown-Kucha\v{r} model, the energy density of the dust fields needs to be chosen negative for some cases. We find numerical solutions   of the background dynamics for both positive and negative dust energy densities in the Brown-Kucha\v{r} model with different initial conditions for a Starobinsky inflationary potential, whereas we considered positive dust energies for the Gaussian dust model only. Our results show that the presence of the dust fields does not prevent the occurrence of the slow-roll inflation as long as the initial energy density of the dust fields is sufficiently small in comparison with that of the scalar field. During the slow-roll phase, the energy density of the dust fields  decays exponentially with the expansion of the universe.

In our numerical investigations, we first set the initial conditions in the pre-inflationary era at the bottom of the potential such that the inflaton rolls up the potential before slow rolling down. We find that the presence of the
dust reference fields does affect the number of inflationary e-foldings by changing the magnitude of Hubble friction in the Klein-Gordon equation (\ref{3a8}). More specifically, when the inflaton climbs up the inflationary potential before the onset of inflation larger Hubble damping due to positive dust energy density leads to a lower value of the field at the onset of inflation and thus results in a smaller number of inflationary e-foldings. In contrast, the negative dust energy density has an opposite effect. In case one chooses initial conditions at the top of the potential such that the inflaton rolls down, due to an increase in the Hubble friction the positive dust energy density increases the number of e-foldings, whereas the negative dust energy density decreases the inflationary e-folds. When the initial dust energy density takes a negative sign, one necessary condition for the slow-roll to take place is that the energy density of the scalar field should be larger than the magnitude of the dust energy density at all times, especially at the turnaround point when the inflaton starts to roll down the potential. Otherwise, instead of the slow-roll, a recollapse of the universe happens, resulting in a big crunch singularity.

The equations of motion of the linear perturbations  are derived from Hamilton's equations  generated by the physical Hamiltonian of the Brown-Kucha\v r  model. It turns out that the final equations of motion for the manifestly gauge-invariant observables in the Gaussian dust model can be simply obtained from (\ref{4b2})-(\ref{4b2a}) when all terms that involve the observables associated with the shift vector $\overline{N}^i$ and its perturbations $\delta N^i$ are removed. The reason for this is that in the Gaussian dust model in the reduced phase space of full GR, we have $N^i=0$ and thus allowing only a vanishing background and vanishing perturbations for the shift vector degrees of freedom. 
By means of a standard SVT decomposition commonly used in  conventional cosmological perturbation theory, the manifestly gauge-invariant linear perturbations in the relational formalism can be decomposed into three scalar modes, two vector modes and two tensor modes, all of which are now physical modes. Given these manifestly gauge-invariant quantities in the scalar sector we chose an independent set of variables and constructed the image of the Mukhanov-Sasaki variable under the observable map in the reduced phase space and derived the corresponding Mukhanov-Sasaki equation in the relational formalism.

We found that in both of the models for this chosen set of variables in the reduced phase space the Mukhanov-Sasaki equation for the analogue of the Mukhanov-Sasaki variable defined in (\ref{4b8}) differs from the conventional one by an additional term $F^{\rm BK/G}_\mathrm{dust}$ that will be absent if no dust reference fields are considered. Moreover, due to the specific combination of the scalar modes used to define the Mukhanov-Sasaki variable, the Mukhanov-Sasaki equation turns out to take the same form in the two dust models. However, several differences between the two dust models are also observed. First, although the perturbation of the dust momentum density $\delta \mathcal E^{\mathrm{dust}}_{\parallelsum}$ is always a constant of motion in the two dust models, the perturbation of the dust energy density $\delta \mathcal E^{\mathrm{dust}}$ is not a constant of motion for a nonvanishing  $\delta \mathcal E^{\mathrm{dust}}_{\parallelsum}$ in the Gaussian dust model. However,  $\delta \mathcal E^{\mathrm{dust}}$ is a constant of motion in the Brown-Kucha\v r dust model due to the extra terms related with $\delta \mathcal E^{\mathrm{dust}}_{\parallelsum}$ in the equations of motion.
As explained above exactly this different behavior in the two dust models is expected in order to ensure that they are consistent with the perturbed constraint algebra.
Secondly,  although the Mukhanov-Sasaki equation takes the same form in the two dust models, the dust correction term also involves the other physical scalar modes which shows that the  Mukhanov-Sasaki equation not in a closed form. From our numerical solutions of the linear perturbations, we find that the Gaussian dust and the Brown-Kucha\v r dust models can have different physical results. As a first step towards understanding the imprints of chosen reference fields in more detail, we estimated the evolution of the background coefficients in front of the linear perturbations involved in $F^{\rm BK/G}_\mathrm{dust}$.  We found that these coefficients decay rapidly during inflation. The form of $F^{\rm BK/G}_\mathrm{dust}$ for our chosen set of variables on the reduced phase space has a rather complicated form. The choice of variables is guided by the aim of comparing the reduced phase space obtained from the dust reference fields with the conventional case where a convenient gauge for the Mukhanov-Sasaki equation is the spatially flat gauge. It will be interesting to investigate whether there are other sets of variables on the reduced phase space that allow to simplify the explicit form of $F^{\rm BK/G}_\mathrm{dust}$ in these other variables. While we do not expect that the decay of background coefficients of these corrections during inflation depends on the chosen set of gauge-invariant variables as long as it involves the Mukhanov-Sasaki variable we used in this work here, such variables can simplify extracting the primordial power spectrum of perturbations in the Brown-Kucha\v{r} and Gaussian dust models and their comparison. The way dust reference fields affect the power spectrum of perturbations for different initial states is an open question which is worthy to be explored in a future work.

\section*{Acknowledgements} 
This work is  supported by the DFG-NSF grants PHY-1912274 and 425333893. K.G. and L.H. thank Louisiana State University for hospitality and support during visits where parts of this work were completed. 


\end{document}